\documentclass[%
preprint,
 amsmath,amssymb,
 aps,
]{revtex4-1}

\usepackage{graphicx} 
\graphicspath{ {Figures/} }
\usepackage{epstopdf}
\usepackage[caption=false]{subfig}

\usepackage{amsmath}
\usepackage{amsfonts}
\usepackage{amssymb}
\usepackage{bm}
\usepackage{hyperref}
\usepackage{amsthm}

\usepackage{color}

\usepackage{graphicx,color}
\usepackage{amsmath}
\usepackage{amssymb}
\usepackage{mathrsfs}
\usepackage[vlined,ruled]{algorithm2e}
\usepackage{url}
\usepackage{color}
\usepackage{algorithmic}
\usepackage{bbm}

\usepackage{booktabs}
\usepackage{array}
\usepackage[table]{xcolor}
\usepackage{yfonts}

\newtheorem{theorem}{Theorem}[section]
\newtheorem{lemma}[theorem]{Lemma}

\newtheorem{assumption}{Assumption}

\newcommand{\map}[3]{#1: #2 \rightarrow #3}

\newcommand{\subscr}[2]{{#1}_{\textup{#2}}}
\newcommand{\supscr}[2]{{#1}^{\textup{#2}}}
\newcommand{\until}[1]{\{1,\dots,#1\}}

\newcommand{\Image}{\operatorname{Im}}

\newcommand{\real}{\mathbb{R}}

\newcommand{\mc}{\mathcal}

\DeclareSymbolFont{bbold}{U}{bbold}{m}{n}
\DeclareSymbolFontAlphabet{\mathbbold}{bbold}

\newcommand\oprocendsymbol{\hbox{$\square$}}
\newcommand\oprocend{\relax\ifmmode\else\unskip\hfill\fi\oprocendsymbol}

\usepackage{tikz}
\usetikzlibrary{matrix}
\usepackage{amsmath}

\begin{document}
\bibliographystyle{naturemag}

\title{Topological Principles of Control in Dynamical Network Systems}
\author{Jason Kim}
\affiliation{Department of Bioengineering, University of Pennsylvania, Philadelphia, PA, 19104}
\author{Jonathan M. Soffer}
\affiliation{Department of Bioengineering, University of Pennsylvania, Philadelphia, PA, 19104}
\author{Ari E. Kahn}
\affiliation{Department of Neuroscience, University of Pennsylvania, Philadelphia, PA, 19104}
\affiliation{U.S. Army Research Laboratory, Aberdeen, MD 21001}
\author{Jean M. Vettel}
\affiliation{Human Research \& Engineering Directorate, U.S. Army Research Laboratory, Aberdeen, MD 21001}
\affiliation{Department of Bioengineering, University of Pennsylvania, Philadelphia, PA, 19104}
\affiliation{Department of Psychological and Brain Sciences, University of California, Santa Barbara, CA, 93106}
\author{Fabio Pasqualetti}
\affiliation{Department of Mechanical Engineering, University of California, Riverside, Riverside, CA, 92521}
\author{Danielle S. Bassett}
\affiliation{Department of Bioengineering, University of Pennsylvania, Philadelphia, PA, 19104}
\affiliation{Department of Electrical and Systems Engineering, University of Pennsylvania, Philadelphia, PA, 19104}
\affiliation{To whom correspondence should be addressed: dsb@seas.upenn.edu}
\date{\today}

\clearpage
\newpage
\begin{abstract}
~\\
~\\
Networked systems display complex patterns of interactions between a large number of components. In physical networks, these interactions often occur along structural connections that link components in a hard-wired connection topology, supporting a variety of system-wide dynamical behaviors such as synchronization and correlated activity. While descriptions of these behaviors are important, they are only a first step towards understanding the relationship between network topology and system behavior, and harnessing that relationship to optimally control the system's function. Here, we use linear network control theory to analytically relate the topology of a subset of structural connections (those linking driver nodes to non-driver nodes) to the minimum energy required to control networked systems. As opposed to the numerical computations of control energy, our accurate closed-form expressions yield general structural features in networks that require significantly more or less energy to control, providing topological principles for the design and modification of network behavior. To illustrate the utility of the mathematics, we apply this approach to high-resolution connectomes recently reconstructed from drosophila, mouse, and human brains. We use these principles to show that connectomes of increasingly complex species are wired to reduce control energy. We then use the analytical expressions we derive to perform targeted manipulation of the brain's control profile by removing single edges in the network, a manipulation that is accessible to current clinical techniques in patients with neurological disorders. Cross-species comparisons suggest an advantage of the human brain in supporting diverse network dynamics with small energetic costs, while remaining unexpectedly robust to perturbations. Generally, our results ground the expectation of a system's dynamical behavior in its network architecture, and directly inspire new directions in network analysis and design via distributed control.
\end{abstract}

\maketitle

\section{Introduction}

Network systems are composed of interconnected units that interact with each other on diverse temporal and spatial scales \cite{newman2010networks}. The exact patterns of interconnections between these units can take on many different forms, and those forms can dictate how the system functions \cite{newman2003structure}. Indeed, specific features of network topology -- such as small-worldness \cite{Watts1998} and modularity \cite{simon1962architecture} -- can give rise to properties like control efficiency \cite{Latora2001,Vragovic2005} and robustness against component failure \cite{Joyce2013} that are highly desirable for both natural and man-made systems. While the relationship between interconnection architecture and dynamics is observed ubiquitously across technological, social, and physical systems, it provides particularly important insights into the functional capabilities of biological systems such as the brain. Here, the topology of interconnection pattterns between neural units is thought to support optimal information processing \cite{bassett2006small,bassett2016small,sporns2016modular}, both at the level of individual cells \cite{bettencourt2007functional} and at the level of meso-scale brain areas \cite{bassett2006adaptive,bullmore2009complex}. 

Despite the observation that network topology and system function are related to one another, there still remains very little understanding of the exact mechanisms driving this relationship \cite{pasqualetti2014controllability}. Gaining such an understanding would have far reaching implications for the analysis, modification, and control of interconnected complex systems \cite{newman2006structure}. This relationship could be exploited for personalized therapeutics \cite{barabasi2011network} that would significantly enhance clinical outcomes for patients by mapping the network topology in musculoskeletal \cite{murphy2016structure}, gene regulatory \cite{conaco2012functionalization,henzler2013staged,norton2016detecting}, and central nervous \cite{bassett2016network} systems. For example, treatments for drug-resistant epilepsy currently range from surgical resection or laser ablation of epileptic tissue to stimulation designed to stem seizure progression. Yet, these interventions are often complicated by the presence of epileptic tissue in areas of the brain that are essential for motor and language function \cite{Nune2015}. An understanding of the specific role of regions and connections in brain networks could inform more targeted and less invasive therapies to make the epileptic state energetically unfavorable to reach, or impossible to maintain \cite{ching2012distributed,khambhati2015dynamic,khambhati2016virtual}. 

Existing paradigms for exploring the mechanisms by which a complex network topology drives observable dynamics come from diverse intellectual fields and are built on varying assumptions. One paradigm stems from the field of nonlinear dynamics, and deals with attractors (states a system naturally tends to) and basins of attraction (regions of initial states that naturally fall into an attractor) \cite{Sprott2015}. This is an effective tool for understanding meso-scale networks that display nonlinear dynamics, and for defining perturbations of state trajectories to force a system to transition from one basin to another \cite{Cornelius2013}. A principal challenge of this approach is that analytical solutions explaining mechanisms of network control remain sparse. An alternative paradigm involves calculating statistical correlations between observed function and structure using graph theoretical metrics such as network communicability \cite{estrada2008communicability}, modularity \cite{newman2006modularity}, or search information \cite{trusina2005communication,rosvall2005searchability,sneppen2005hide}. This approach has been used in the context of brain networks to predict -- in a statistical sense -- patterns of activity from patterns of wiring \cite{goni2014resting}, the impact of focal lesions on distant brain areas following stroke \cite{crofts2011network}, and the role of modules in facilitating adaptive functions such as learning \cite{Bassett2011,mantzaris2013dynamic,bassett2013task,bassett2014cross,bassett2015learning}. The principal challenge of this approach is that graph statistics themselves do not constitute mechanisms in a philosophical \cite{craver2005beyond} or mathematical sense \cite{bassett2016mattar}.

A promising paradigm that meets these challenges is linear network control theory \cite{kalman1963mathematical,kalman1963controllability,lin1974structural}, which assumes that the dynamics of a system are state-dependent, and that they are linear with respect to the state.  In this framework, the state of a system at a given time is a function of the previous state, the structural network linking units of that system, and any input to the system provided in the form of control energy (see next section for explicit mathematical definitions). From this paradigm arises the possibility of identifying driver nodes in the system \cite{liu2011controllability,campbell2015topological,ruths2014control}: units that have the potential to influence the system along diverse control strategies. Also, within this framework we can compute optimal inputs that move the system from one state to another with minimal cost. This formulation has been particularly useful in understanding a variety of networked systems, including the human brain where control points facilitate diverse cognitive strategies \cite{Gu2015,gu2016optimal}, facilitate efficient activation and deactivation patterns during endogeneous activity \cite{Betzel2016}, and inform optimal targets for brain stimulation to alter neural activity \cite{muldoon2016stimulation}. 

While the identification of control points and optimal trajectories is computationally tractable, basic intuitions about the network properties that enhance control -- either locally around a given node or globally throughout the whole system -- have remained elusive. In this study, we sought to identify the key topological features that determine network controllability, and use these principles to understand and modify the complex connectivity of meso-scale brain networks in a dynamically meaningful way. To reach these goals, we formulate a linear control problem that examines a subset of a network's edges: the connections that link driver nodes to non-driver nodes. Using this bipartite subgraph of the entire network, we explore the mathematical theory of fundamental driver$\to$non-driver dynamics to the extent of explicitly understanding the role of every node and edge in a networked system, and modifying those objects to alter control energy in an exactly predictable manner. We show that the intuitions and solutions for the control of the bipartite subgraph provide excellent estimates of the control of the full network. Our results include analytical derivations of expressions relating a network's minimum control energy to its connection topology, along with an intuitive geometric representation to visualize this relationship. While our mathematical contributions are applicable to any complex network system whose dynamics can be approximated by a linear model, we illustrate the utility of the formulation in the context of brain networks estimated from the mouse brain (made publically available by the Allen Brain Institute) \cite{oh2014mesoscale,rubinov2015wiring}, the drosophila brain \cite{shih2015connectomics}, and the human brain (Fig.~\ref{fig:intro}d--f). Specifically, we use the analytical expressions to (i) understand key patterns and principles of connectivity that determine a network's control profile, (ii) describe the implications of the connectivity of brain networks on their control profiles, and (ii) explicitly modify the control properties of the brain by performing energetically favorable edge deletions, thereby informing potential clinical interventions. Together, these results offer fundamental insights into key patterns of connections between brain regions that directly impact their minimum control energy, providing a link between the structure and function of neural systems.

\section{Network Topology and Controllability}

\begin{figure}[h!]
	\centering
	\includegraphics[width=.9\columnwidth]{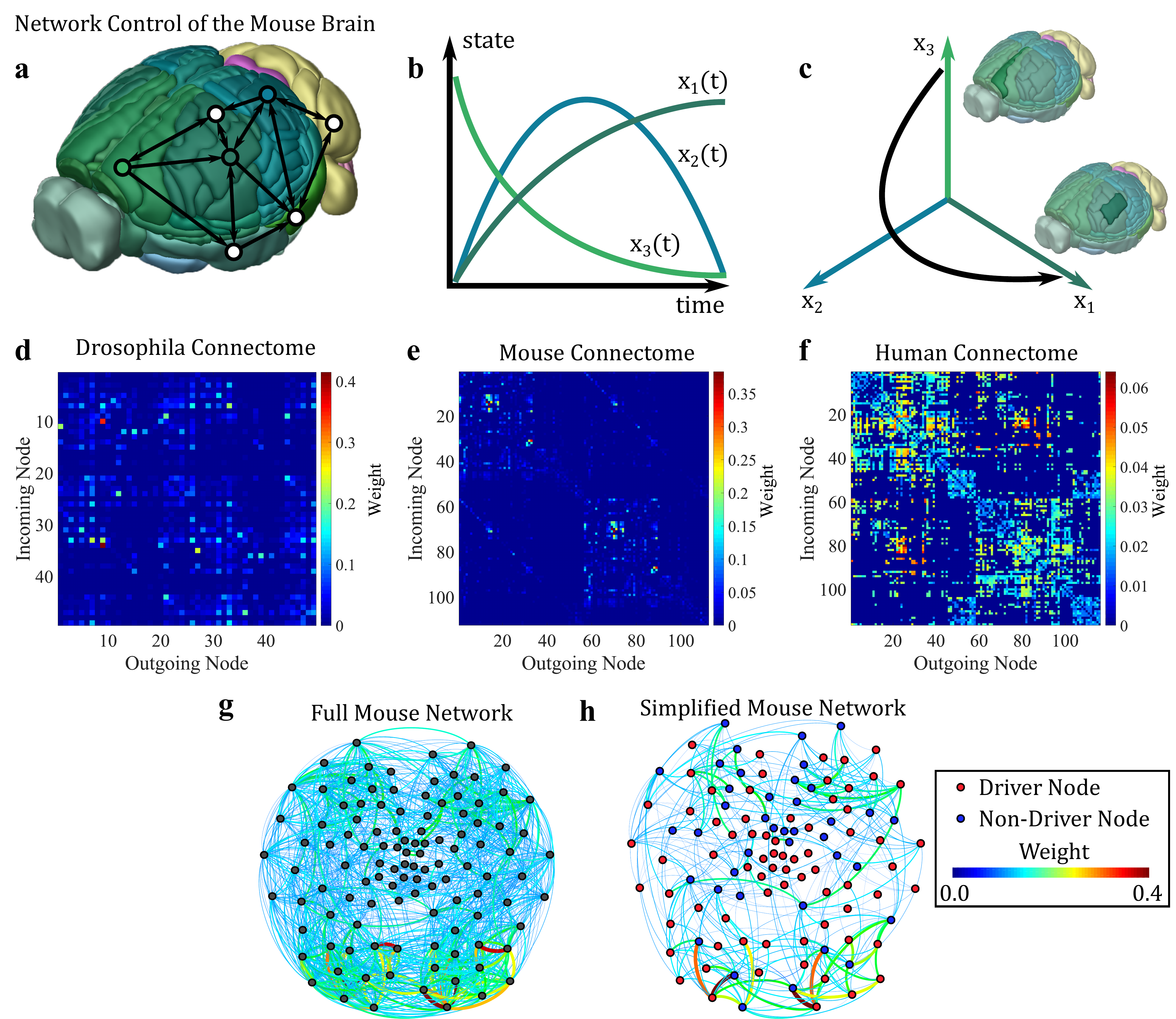}
	\caption{\textbf{Network Control of the Drosophila, Mouse, and Human Connectomes.} (\textbf{a}) A representation of the mouse brain via the Allen Mouse Brain Atlas, with a superimposed simplified network. Each brain region is represented as a vertex, and the connections between regions are represented as directed edges. (\textbf{b}) Example trajectories of state over time for three brain regions, where the state represents the level of activity in each region. (\textbf{c}) A state-space representation of activity on the mouse connectome over time, where each point on the black line represents the brain state at a point in time. (\textbf{d}) Connectomes represented as $n \times n$ adjacency matrices where each $i,j$th element of the adjacency matrix represents the strength of the connection from node $j$ to node $i$ for a drosophila, (\textbf{e}) mouse, and (\textbf{f}) human. (\textbf{g}) The mouse connectome represented as a graph with vertices as brain regions, and edges colored by their weight, or the magnitude of the relevant element of the adjacency matrix. (\textbf{h}) Simplified graph representation: a bipartite subgraph containing edges linking driver vertices (red) to non-driver vertices (blue).}
	\label{fig:intro}
\end{figure}

We are particularly interested in understanding how a network's specific topological features (edge connections, edge weights, weight distributions) affect the energy required to control the network. We are also interested in how that same topology facilitates or inhibits certain states from being reached. To this end, we develop our analysis tools based on a simplified network model, which effectively reveals hidden relations between network topology and control energy. Finally, we validate the predictive power of our results on network models representing the mouse, drosophila, and human brains.

We consider a network represented by the directed graph $\mc G = (\mc V, \mc E)$, where $\mc V = \until{n}$ and $\mc E \subseteq \mc V \times \mc V$ are the sets of network vertices and edges, respectively. Let $a_{ij} \in \real$ be the weight associated with the edge $(i,j) \in \mc E$, and let $A = [a_{ij}]$ be the weighted adjacency matrix of $\mc G$. We associate a real value (\emph{state}) with each node, collect the nodes' states into a vector (\emph{network state}), and define the map $\map{\bm{x}}{\real_{\ge 0}}{\real^n}$ to describe the evolution (\emph{dynamics}) of the network state over time (Fig.~\ref{fig:intro}a--c). We assume that a subset of $N$ nodes, called drivers, is independently
manipulated by external controls and, without loss of generality, we reorder the network nodes such that the $N$ drivers come first. Thus, the network dynamics with controlled drivers read as
\begin{align}\label{eq: control dynamics L}
\begin{bmatrix}
\subscr{\dot{\bm{x}}}{d} \\ 
\subscr{\dot{\bm{x}}}{nd}
\end{bmatrix}
=
\begin{bmatrix}
A_{11} & A_{12}\\
A_{21} & A_{22}
\end{bmatrix}
\begin{bmatrix}
\subscr{\bm{x}}{d} \\ 
\subscr{\bm{x}}{nd}
\end{bmatrix}
+
\begin{bmatrix}
I_N \\ 0
\end{bmatrix}
\bm{u},
\end{align}
where $\subscr{\bm{x}}{d}$ and $\subscr{\bm{x}}{nd}$ are the state vectors of the driver and non-driver nodes, $A_{11} \in \real^{N \times N}$, $M = n - N$, $A_{12} \in \real^{N \times M}$, $A_{21} \in \real^{M \times N}$, $A_{22} \in \real^{M \times M}$, $I_N$ is the $N$-dimensional identity matrix, and $\map{\bm{u}}{\real_{\ge 0}}{\real^N}$ is the control input.

We refer to the network as \emph{controllable} at time $T \in \real_{\ge 0}$ if, for any pair of states $\subscr{\bm{x}}{d}^*$ and $\subscr{\bm{x}}{nd}^*$, there exists a control input $u$ for the
dynamics Eq.~\eqref{eq: control dynamics L} such that $\subscr{\bm{x}}{d}(T) = \subscr{\bm{x}}{d}^*$ and $\subscr{\bm{x}}{nd}(T) = \subscr{\bm{x}}{nd}^*$. For a detailed discussion and rigorous conditions for
the controllability of a system with linear dynamics, see \cite{TK:80}. Finally, we define the energy of $\bm{u}$ as 
\begin{align*}
\textup{E} (\bm{u}) = \sum_{i = 1}^N \underbrace{\int_0^T  u_i
	(t)^2 dt}_{\textup{E}_i} ,
\end{align*}
where $u_i$ is the $i$-th component of $\bm{u}$. The energy of $u_i$ can be thought of as a quadratic cost that penalizes large control inputs to drive the system. In what follows, we characterize how a network's topology determines the control energy needed for a given control task.

We approach this problem by approximating the interactions between brain regions as linear, time invariant dynamics, where a stronger structural connection between two regions represents a stronger dynamic interaction (for empirical motivation for this approximation, see \cite{Gu2015,galan2008network,honey2009predicting}). From these dynamics, we identify physically meaningful interpretations of the underlying mathematical features, which determine the controllability of the simplified network only containing connections from drivers $\to$ non-drivers (Fig.~\ref{fig:intro}h). To justify this approximation, we show that the simplified network well approximates the control dynamics along the full network (Fig.~\ref{fig:intro}g) for a wide range of model parameters. From this simplification, we derive a closed-form expression of the minimum control energy that shows that the similarity in these driver $\to$ non-driver connections scales the energy required to control the network. Given this scaling, we show that dissimilarly connected regions are easiest to control, while similarly connected regions are most difficult to control. Finally, we use these principles to gain insight into the relationship between the structural connectivity and function of brain networks, and make a few strategic topological modifications to affect a profound reduction in control energy. We conclude by discussing the utility of these insights in informing interventions to modulate brain dynamics.

\section{Results}

\subsection{Control energy is well predicted by direct connections between driver and non-driver nodes}

We seek an accurate, tractable relationship between the topology of a network and the energy required to drive the network from one state to another. To find this relationship, we look to classic results in the mathematical theory of systems and control \cite{TK:80}, where the spectral properties of the \textit{reachability Gramian} $W_R(0,T) = \int_0^T e^{At}BB^Te^{A^Tt}$ quantify the
minimum amount of energy (Section~\ref{sec: validation}) to control the network Eq.~\eqref{eq: control dynamics L}. Explicit formulas and bounds for the eigenvalues of the Gramian are intricate and hard to derive, and approximate formulations are often preferred. 

\begin{figure}[h]
	\centering
	\includegraphics[width=1\columnwidth]{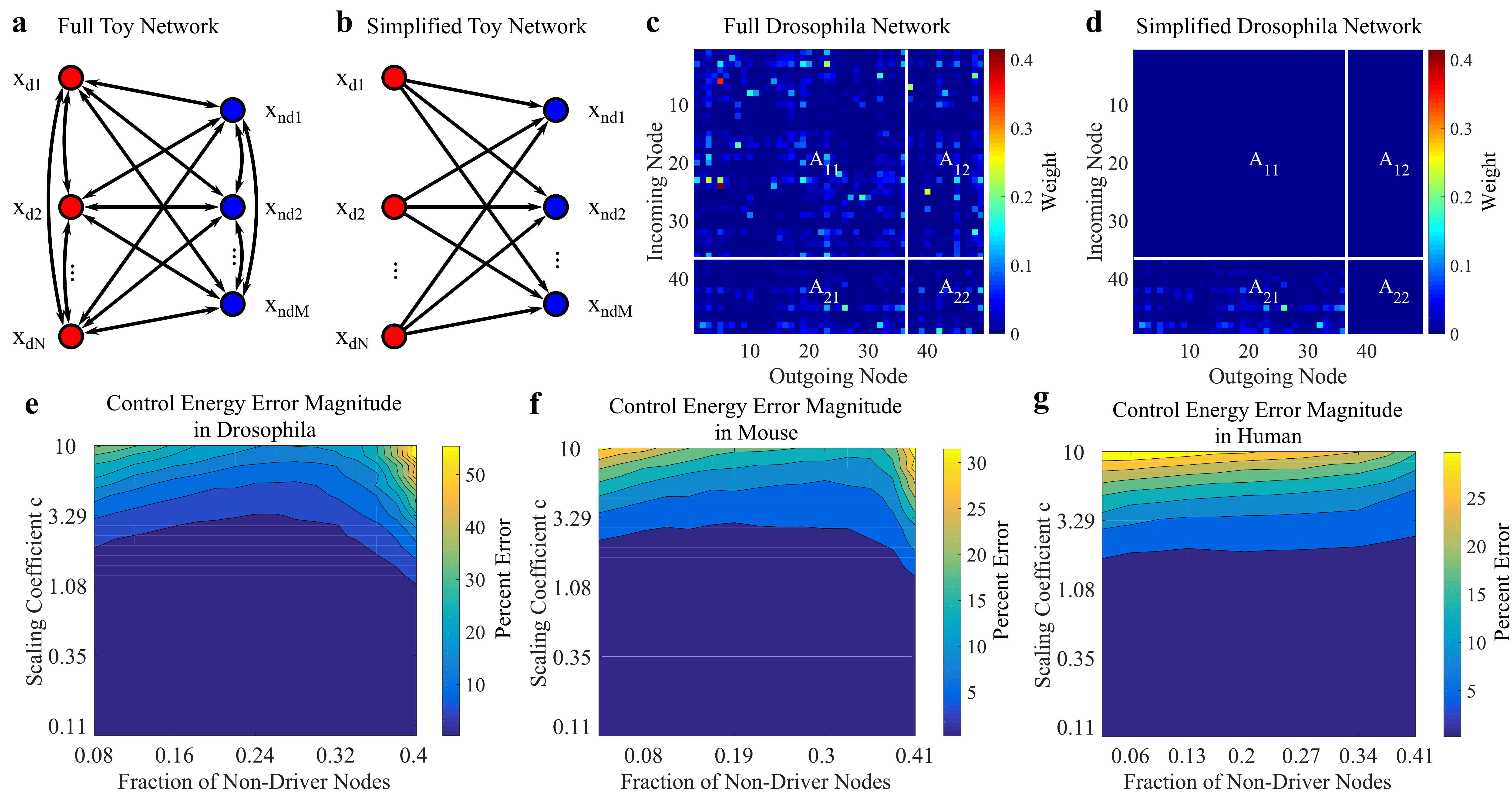}
	\caption{\textbf{The Simplified Network Representation Offers a Reasonable Prediction for the Full Network's Control Energy.} (\textbf{a}) Graphical representation of a \emph{non-simplified} network of $N$ drivers (red) and $M$ non-drivers (blue), with directed connections between all nodes present. (\textbf{b}) Graphical representation of a \emph{simplified} first-order network only containing first-order connections from drivers $\to$ non-drivers. (\textbf{c}) As an example, we show the adjacency matrix for the drosophila connectome segmented into driver $\to$ driver $\subscr{A}{11}$, driver $\to$ non-driver $\subscr{A}{21}$, non-driver $\to$ driver $\subscr{A}{12}$, and non-driver $\to$ non-driver $\subscr{A}{22}$ sections for a non-simplified network as per Eq. \eqref{eq: control dynamics L}, with randomly designated driver and non-driver nodes, and (\textbf{d}) the corresponding simplified network as per Eq. \eqref{eq: energy approximation}. (\textbf{e}) Percent error contour plots of the total control energy for simplified \emph{versus} non-simplified networks as a function of the fraction of non-driver nodes and matrix scale given by coefficient $c$ times the adjacency matrix $A$ normalized by its largest eigenvalue. The median error magnitude for 1000 iterations per combination is shown. Each contour represents a 5\% interval for the (\textbf{e}) drosophila, (\textbf{f}) mouse, and (\textbf{g}) human connectome.}
	\label{fig:phase}
\end{figure}

To make progress on this problem, we begin by calling the network involving edges between all nodes a \emph{non-simplified network} (Fig.~\ref{fig:phase}a), and the network involving only the edges from the driver to the non-driver nodes a \emph{simplified, first-order network} (Fig.~\ref{fig:phase}b). We then derive an accurate approximation of the minimum control energy (Lemma~\ref{lemma: minimum control law L} - \ref{lemma: total control energy L}) by assuming that $\subscr{\bm{x}}{d}(0) = 0$, $\subscr{\bm{x}}{nd}(0) = 0$ (Assumption~\ref{assumption: initial states}), and $A_{11} = 0$, $A_{12} = 0$, and $A_{22} = 0$ (Assumption~\ref{assumption: network edges}) in Eq.~\eqref{eq:
  control dynamics L}, which reads as
\begin{align}\label{eq: energy approximation}
  \textup{E}(\bm{u}) = 12(\subscr{\bm{x}}{nd}^* -
  \frac{1}{2}\subscr{A}{21}\subscr{\bm{x}}{d}^*)^T(
  \subscr{A}{21}\subscr{A}{21}^T)^{-1}( \subscr{\bm{x}}{nd}^* -
  \frac{1}{2}\subscr{A}{21} \subscr{\bm{x}}{d}^*) + \subscr{\bm{x}}{d}^{*T}
  \subscr{\bm{x}}{d}^* ,
\end{align}
where $\subscr{\bm{x}}{d}^*$ and $\subscr{\bm{x}}{nd}^*$ are the desired final states of the driver and non-driver nodes, respectively. We make Assumption~\ref{assumption: initial states} because we are interested in the \emph{change} in brain state through control, and consider initial conditions $\subscr{\bm{x}}{d}(0) = 0$, $\subscr{\bm{x}}{nd}(0) = 0$ to be a neutral baseline.  

Because the expression Eq.~\eqref{eq: energy approximation} involves only the edges in the simplified network from the driver to the non-driver nodes, that is, the matrix $A_{21}$, we say that Eq.~\eqref{eq: energy approximation} is a first-order approximation to the minimum control energy of the non-simplified network Eq.~\eqref{eq: control dynamics L}. One topological feature that impacts the accuracy of the first-order energy approximation is the fraction of nodes that are selected as non-drivers: the drivers of a simplified first-order network (Fig.~\ref{fig:phase}b) can only control the non-drivers through the direct driver $\to$ non-driver connections, while a non-simplified network (Fig.~\ref{fig:phase}a) can additionally use indirect non-driver $\to$ non-driver connections. In a network with few non-drivers, there are relatively few indirect connections, so most of the control dynamics rely upon the direct connections; in a network with many non-drivers, there are more indirect connections for the drivers to utilize, making the first-order control energy Eq.~\eqref{eq: energy approximation} a worse approximation. A second topological feature that affects the accuracy of the first-order approximation is the relative scaling of the adjacency matrix $A$, given by a constant $c$ times every element in $A$. For matrix $A$ multiplied by some scaling coefficient $c$, the natural dynamics along the network without control obey $\dot{\bm{x}} = cA\bm{x}$, which implies $\ddot{\bm{x}} = cA\dot{\bm{x}} = c^2A^2\bm{x}$, and $\bm{x}^{[k]} = c^kA^k\bm{x}$. In simplified networks, $A^k = 0$ for $k > 1$. Hence, for small $c$, the natural dynamics along the simplified and non-simplified networks are similar. The larger the scaling coefficient $c$, the more the non-simplified network dynamics deviate from the simplified network dynamics. 

We analyzed the accuracy of this first-order approximation, and its dependence on the scaling coefficient $c$ and fraction of non-driver nodes, in brain networks of several species: the empirical meso-scale mouse connectome consisting of 112 interconnected brain regions from the Allen Brain Institute, a drosophila connectome consisting of 49 interconnected brain regions \cite{shih2015connectomics}, and a set of human connectomes consisting of 116 brain regions interconnected by white matter tracts estimated using a 705-direction diffusion imaging scan over 55 minutes (for empirical details regarding connectivity estimates, see Methods; for a conceptual schematic of the full and simplified Drosophila connectome, see Fig.~\ref{fig:phase}c--d). We normalized each connectome by the magnitude of its largest eigenvalue \cite{Betzel2016}, and selected a range of scaling coefficients $c$ and fractions of non-driver nodes $d$. For each combination of scaling coefficient $c_i$ and fraction of non-driver nodes $d_j$, we selected 1000 random permutations of drivers and non-drivers, and computed the minimum energy required to drive the simplified and non-simplified networks from initial states $\subscr{\bm{x}}{d} = 0$, $\subscr{\bm{x}}{nd} = 0$ to random final states $\subscr{\bm{x}}{nd}^* \in (-1,1)^M,\subscr{\bm{x}}{d}^* \in (-1,1)^N$. We then calculated the median magnitude of the percent error for the control energies between the simplified and non-simplified networks (Fig.~\ref{fig:phase}e--g for drosophila, mouse, and human, respectively).

We observed that the percent error magnitude between the control energies along the non-simplified \emph{versus} simplified networks was similar across species. In general, the error remained below approximately 5\% for scaling coefficients $c < 1.5$, and fraction of non-driver nodes $d < 0.4$, and below 10\% for $c < 3.3$, and $d < 0.39$ (Fig.~\ref{fig:phase}e--g), confirming that the first-order energy approximation is accurate within a range of scaling coefficients and non-driver fractions for these empirical connectomes. More generally, the accurate, closed-form expression relating a network's topology and control energy can be used to extract the underlying topological features that determine controllability. For the remainder of this paper, we will use the same connectomes (drosophila, mouse, and human) at a scaling coefficient of $c=1$, and non-driver fraction $d\leq0.4$, to ensure generalizability of our findings to the non-simplified versions of these same networks. For any examples requiring a specific fraction of non-drivers, we will use a fraction of 0.2, corresponding to $M=10$ in drosophila, $M=22$ in mouse, and $M=23$ in human connectomes.

\subsection{Determinant of the driver-to-non-driver connection matrix scales the control energy}

\begin{figure}[h]
	\centering
	\includegraphics[width=1.0\columnwidth]{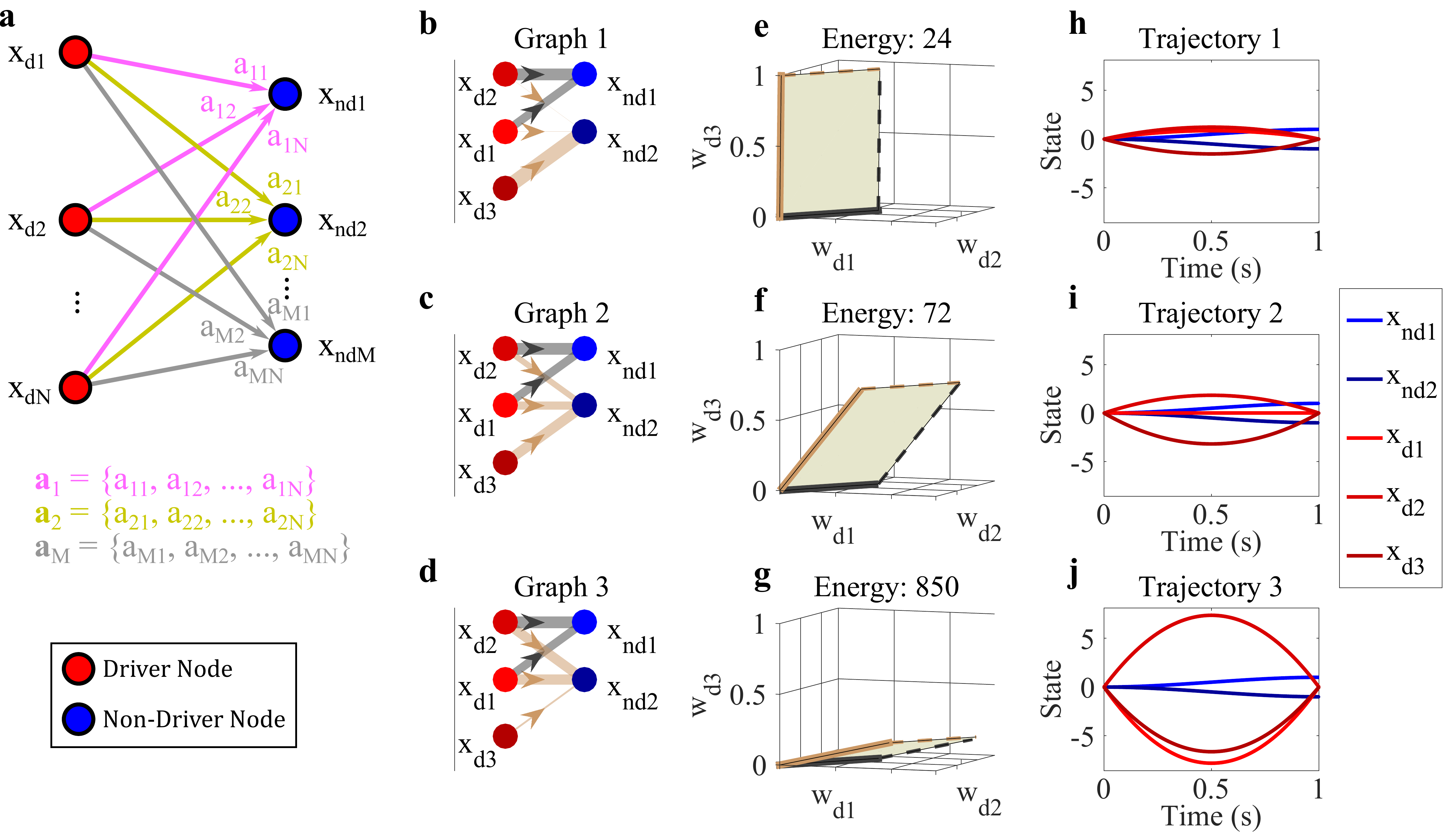}
	\caption{\textbf{Geometric Interpretation of Simplified, First-Order Networks with Corresponding Control Energies and Trajectories} (\textbf{a}) Graph representation of a simplified first-order network containing connections from $N$ driver nodes in red to  $M$ non-driver nodes in blue. The edges connecting all driver nodes to the $i$-th non-driver corresponding to the $i$-th row of $\subscr{A}{21}$ are shown in different colors. (\textbf{b}) Graph representation of a network with driver nodes in red, non-driver nodes in blue, weight distribution into non-driver 1 in gray, and weight distribution into non-driver 2 in tan, for dissimilarly distributed weights, (\textbf{c}) for somewhat similarly distributed weights, and (\textbf{d}) for very similarly distributed weights. (\textbf{e}) Geometric representation of the parallelotope formed by the 2 vectors of weight distributions into non-drivers 1 and 2, with the volume shaded in beige for dissimilarly distributed weights, (\textbf{f}) for somewhat similarly distributed weights, and (\textbf{g}) for very similarly distributed weights. (\textbf{h}) The states of all network nodes over time for dissimilarly distributed weights, (\textbf{i}) for somewhat similarly distributed weights, and (\textbf{j}) for very similarly distributed weights.}
	\label{fig:weight_vector}
\end{figure}

After deriving a closed-form approximation for the minimal total energy required to drive a network from one state to another, we next sought to provide a physical interpretation of the mathematical features that drive the control energy. First, we let $Q = \subscr{A}{21}\subscr{A}{21}^T$, and notice that Eq.~\eqref{eq: energy approximation} can be rewritten as 
\begin{align}\label{eq: energy determinant}
  \textup{E} (\bm{u}) = 12 \frac{\bm{v}_1^T\text{adj}(Q)\bm{v}_1}{\det(Q)} + \bm{v}_2^T\bm{v}_2 ,
\end{align}
where $\bm{v}_1 = \subscr{\bm{x}}{nd}^*-\frac{1}{2}\subscr{A}{21}\subscr{\bm{x}}{d}^*$ and $\bm{v}_2 = \subscr{\bm{x}}{d}^*$, and $\text{adj}(Q)$ is the adjoint matrix of $Q$. We notice that, independently of the vectors $v_1$ and $v_2$, the determinant of $Q$ acts as a scaling factor for the total energy. This insight is useful because of the intricate geometric interpretation of a Gram matrix determinant. Specifically, let $\bm{a}_i \in \real^{1 \times N}$ be the $i$-th row of $\subscr{A}{21}$ (which we will call the \emph{weight vector}), representing the connections from all $N$ drivers to the $i$-th non-driver node (Fig.~\ref{fig:weight_vector}a). Then, the determinant of the gram matrix $Q$ is equal to the squared volume of the parallelotope formed by all $\bm{a}_i$.

To gain an intuition for these results, we show a simple system with 3 drivers and 2 non-drivers with varying network topologies in Fig.~\ref{fig:weight_vector}b--d, their corresponding geometric parallelotopes in Fig.~\ref{fig:weight_vector}e--g, with $\bm{a}_1$ as the vector of gray-colored weighted connections into $\subscr{\bm{x}}{nd1}$, and $\bm{a}_2$ as the tan-colored connections into $\subscr{\bm{x}}{nd2}$. The control task has initial states $\subscr{\bm{x}}{d}(0) = 0$, $\subscr{\bm{x}}{nd}(0) = 0$, and final states $\subscr{\bm{x}}{d}^* = 0$, $\subscr{x}{nd1}^*=1, \subscr{x}{nd2}^*=-1$. These final states hard-code the empirical observation that the most common dynamics of intrinsic activity patterns in large-scale human brain networks are anti-correlated activation states \cite{chen2016,Medaglia2015a,Reddy2017}, often referred to task-positive and task-negative \cite{fox2005}. We see that as the total area of the parallelogram shrinks from Fig.~\ref{fig:weight_vector}e--g, the total control energy to move the non-drivers increases in Fig.~\ref{fig:weight_vector}h--j.

Intuitively, if two non-drivers $\subscr{x}{nd1}, \subscr{x}{nd2}$ are very similarly connected to the drivers, it is difficult to drive one of them independently from the other. Geometrically we show in Fig.~\ref{fig:weight_vector} how similarity between two driver $\to$ non-driver connections $\bm{a}_1, \bm{a}_2$ decreases the volume of the corresponding parallelotope, thereby decreasing the determinant in Eq.~\eqref{eq: energy determinant} and inversely scaling the control energy. We see that this relationship between the determinant and similarity of weight vectors $\bm{a}_i$ persists for any number of drivers and non-drivers where the first-order network is a good approximation. We conclude that the similarity in distribution of weights directed into non-driver nodes scales the control energy through the determinant of $Q$, where more similarly connected non-drivers require more energy to control differentially. This relationship is significant because we can now analyze and modify the connectivity of a network, knowing the topological features that determine its control.

\subsection{Using connection topology to identify energetically favorable control nodes}

In the previous section, we derived an approximate relationship between a network's topology and its minimum control energy, and showed that the similarity in driver to non-drivers connections $\bm{a}_i$ changed the determinant of the gram matrix $Q$, thereby scaling the control energy. Here, we further explore this idea of ``similarity" between connections $\bm{a}_i$, in order to quantify the impact of each individual non-driver on the control energy. With this knowledge, we can begin asking questions about the most or least controllable regions in a network, and how to modify specific connections in a network to improve controllability.

\noindent \textbf{C.1. Derivation of the Main Topological Contributors to Control Energy.} Our analysis is rooted in the intuition that the edge weights $\bm{a}_i$ that maximize the parallelotope volume, thereby facilitating network control, are large in magnitude and orthogonal to each other.
Let $\lambda_i$ and $\bm{e}_i$ be the eigenvalues and eigenvectors of the matrix $Q$ in Eq.~\eqref{eq: energy determinant}. We derive in Lemma~\ref{lemma: gram decomp L}  the equivalent, alternative control energy expression written as
\begin{align}
\textup{E} (\bm{u}) = 12\left( \frac{\sum_{i=1}^M w_i c_i^2}{
	\sum_{i=1}^M w_i} \right) 
\left( \sum_{k=1}^M{\frac{1}{\|\bm{a}_k\|^2
		\sin( \theta_k)^2}}\right) + \bm{v}_2^T\bm{v}_2,
\label{eq: energy_segregation}
\end{align}
\noindent where $w_i = \prod_{j\neq i}^M \lambda_j$, $c_i = \bm{e}_i^T \bm{v}_1$, and $\theta_k$ is the angle formed between $\bm{a}_k$ and the parallelotope formed by $\bm{a}_{j\neq k}$. For $N$ drivers and $M$ non-drivers, we can visualize the $M$ weight vectors $\bm{a}_k$ as forming a parallelotope in an $N$-dimensional space. The variable $\theta_k$ then represents the angle formed between $\bm{a}_k$ and the paralellotope formed by the remaining $M-1$ vectors $\bm{a}_{j\neq k}$. An example with $N=3, M=2$ is shown in Fig.~\ref{fig:weight_vector}e--g, where $\theta_1 = \theta_2$ is the angle between the tan and gray vectors. 

Here, we have segregated the control energy into a task-based ($\frac{\sum_{i=1}^M w_i c_i^2}{
\sum_{i=1}^M w_i}$) and topology-based ($\sum_{k=1}^M{\frac{1}{\|\bm{a}_k\|^2 \sin( \theta_k)^2}}$) term. The task-based term is a weighted average of eigenvalue products $w_i$, weighted by the eigenvector composition $c_i$ of a specific task $\bm{v}_1$. The topology-based term is a sum of elements, where each element is an inverse squared function of the magnitude and angle of each weight vector. This segregation allows us to analyze the topology separate from the specific control task, and shows that each non-driver additively contributes to the total control energy. We note that for a non-driver $i$, this contribution is smallest when $\|\bm{a}_i\|$ and $\theta_i$ are large.

\begin{figure}
	\centering
	\includegraphics[width=1.0\columnwidth]{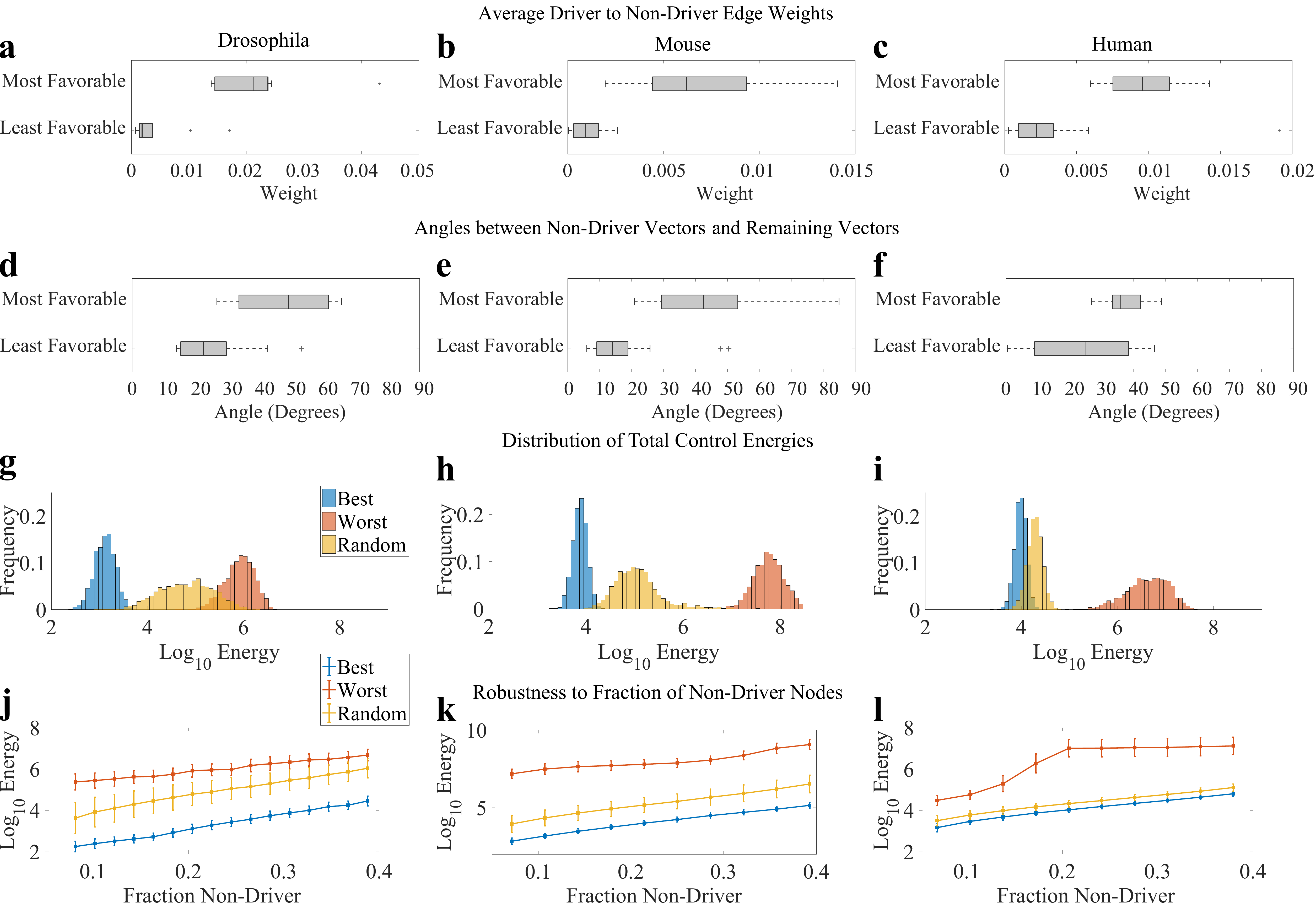}
	\caption{\textbf{Topological Characteristics and Energetic Performance of Networks with Energetically Favorable and Unfavorable Topologies.} (\textbf{a}) Boxplot of the average driver to non-driver edge weights for each non-driver of the energetically most and least favorable networks in the drosophila, (\textbf{b}) mouse, and (\textbf{c}) human connectomes, for a non-driver fraction of 0.2. (\textbf{d}) Boxplot of angles formed by the driver to non-driver connections for each non-driver $i$ and the parallelotope formed by the remaining $M-1$ non-drivers given in Eq.~\eqref{eq: energy_segregation} for the energetically most and least favorable networks in the drosophila, (\textbf{e}) mouse, and (\textbf{f}) human connectomes. (\textbf{g}) Distribution of total control energies along the energetically most favorable, least favorable, and random networks at a non-driver fraction of 0.2 for 2000 control tasks, with initial states $\subscr{\bm{x}}{nd}(0) = 0$, $\subscr{\bm{x}}{d}(0) = 0$, and random final states $\subscr{\bm{x}}{nd}^* \in (-1,1)^M,\subscr{\bm{x}}{d}^* \in (-1,1)^N$ for the drosophila, (\textbf{h}) mouse, and (\textbf{i}) human connectomes. (\textbf{j}) Mean and standard deviations of the base-10 log of the total control energies across various non-driver fractions for the 2000 control tasks along the energetically most favorable, least favorable, and random networks for the drosophila, (\textbf{k}) mouse, and (\textbf{l}) human connectomes.}
	\label{fig:ind_contribution}
\end{figure}

\noindent \textbf{C.2. Most and least energetically favorable driver-nondriver sets in brain connectomes.} To support this discussion, we used the expression Eq.~\eqref{eq: energy_segregation} that segregated the control task from the topology in tandem with a greedy algorithm to find the sets of $M$ non-drivers that minimized and maximized this topology term. We first calculated the topology term for all permutations of 4 non-driver regions, and found the sets of 4 regions that minimized and maximized the topology-based term. Then, we iteratively appended individual regions that minimized and maximized the term until we reached $M$ non-drivers. We defined the \emph{most} and \emph{least energetically favorable} networks to be the selections of $N$ driver and $M$ non-driver nodes that minimize and maximize this topology term, respectively.

As an example using a non-driver fraction of 0.2, we show the distribution of magnitudes $\|\bm{a}_i\|$ of each driver to non-driver weight vector in drosophila, mouse, and human for both the energetically most and least favorable networks (Fig.~\ref{fig:ind_contribution}a--c, respectively). We observe that the energetically least favorable networks have significantly weaker driver to non-driver connections than the energetically most favorable networks via a two-sample $t$-test between the most and least favorable networks in the drosophila ($t(8)=7.19$, $p=1.08\times 10^{-6}$), mouse ($t(20)=7.20$, $p=7.39 \times 10^{-9}$), and human ($t(21)=8.22$, $p=1.93 \times 10^{-10}$).  We also show the corresponding distributions of angles between each driver to non-driver weight vector $\bm{a}_i$ and the parallelotope formed by the remaining $M-1$ non-drivers given by $\theta_i$ in Eq.~\eqref{eq: energy_segregation} in the drosophila, mouse, and human connectomes (Fig.~\ref{fig:ind_contribution}d--f, respectively). We observe that the angles $\theta_i$ in the energetically least favorable networks are significantly smaller than the angles in the energetically most favorable networks via a two-sample $t$-test between the most and least favorable networks in the drosophila ($t(8)=3.67$, $p=1.80 \times 10^{-3}$), mouse ($t(20)=6.04$, $p=3.43 \times 10^{-7}$), and human ($t(21)=3.71$, $p=5.72 \times 10^{-4}$).

Next, we demonstrate the utility and robustness of these topological features for network control by computing the minimum control energy along the non-simplified network using the driver and non-driver designations from the simplified network in Eq.~\eqref{eq: energy_segregation} for a range of non-driver fractions. For each $M$, we calculated the $\log_{10}$ energy from 2000 random control tasks with initial states $\subscr{\bm{x}}{nd}(0) = 0$, $\subscr{\bm{x}}{d}(0) = 0$, and final states $\subscr{\bm{x}}{nd}^* \in (-1,1)^M, \subscr{\bm{x}}{d}^* \in (-1,1)^N$ on the most energetically favorable network, the least energetically favorable network, and a set of randomly chosen $N$ drivers and $M$ non-drivers. We show the control energies across the tasks for the drosophila, mouse, and human connectomes (see Fig.~\ref{fig:ind_contribution}g--i, respectively). We also show the means and standard deviations of the control energy across the random tasks for each $M$ (see Fig.~\ref{fig:ind_contribution}j--l). As can be seen across all three species, the most energetically favorable networks require around 0.5--1 order of magnitude less control energy than the random networks, and 2.5--4 orders of magnitude less control energy than the least energetically favorable networks. This difference indicates an energetic advantage for some configurations of drivers and non-drivers over others.

\subsection{Brain Networks of Increasingly Complex Species have More Energetically Favorable Topological Relationships}

Given the relationship between a network's topology and minimum control energy in Eq.~\eqref{eq: energy_segregation}, we seek to understand if brain networks are organized along energetically favorable principles. Previously, we showed that the control energy was inversely proportional to the squared product of the magnitudes $\|\bm{a}_k\|$ and $\sin(\theta_k)$ of the driver $\to$ non-driver connections. Here, we show that brain networks of increasingly complex species more effectively balance these two topological features to yield robust and energetically favorable connectivities.

Fundamentally, we are asking how well a network's specific set of topological components $\|\bm{a}_k\|$ and $\sin(\theta_k)$ combine to minimize the topology dependent energy term $\sum_{k=1}^M{\frac{1}{\|\bm{a}_k\|^2 \sin( \theta_k)^2}}$. In networks that are not designed along these energetic principles, we expect to see no particular relationship between $\|\bm{a}_k\|$ and $\sin(\theta_k)$. In networks that minimize the topology dependent energy term, we expect to see a compensatory effect, where non-drivers with small angles have large magnitudes, and non-drivers with small magnitudes have large angles. While it may seem intuitive that pairing large magnitudes with large angles would yield lower energies, this strategy also requires smaller magnitudes to pair with smaller angles, yielding disproportionately large energy contributions.

\begin{figure}
	\centering
	\includegraphics[width=0.8\columnwidth]{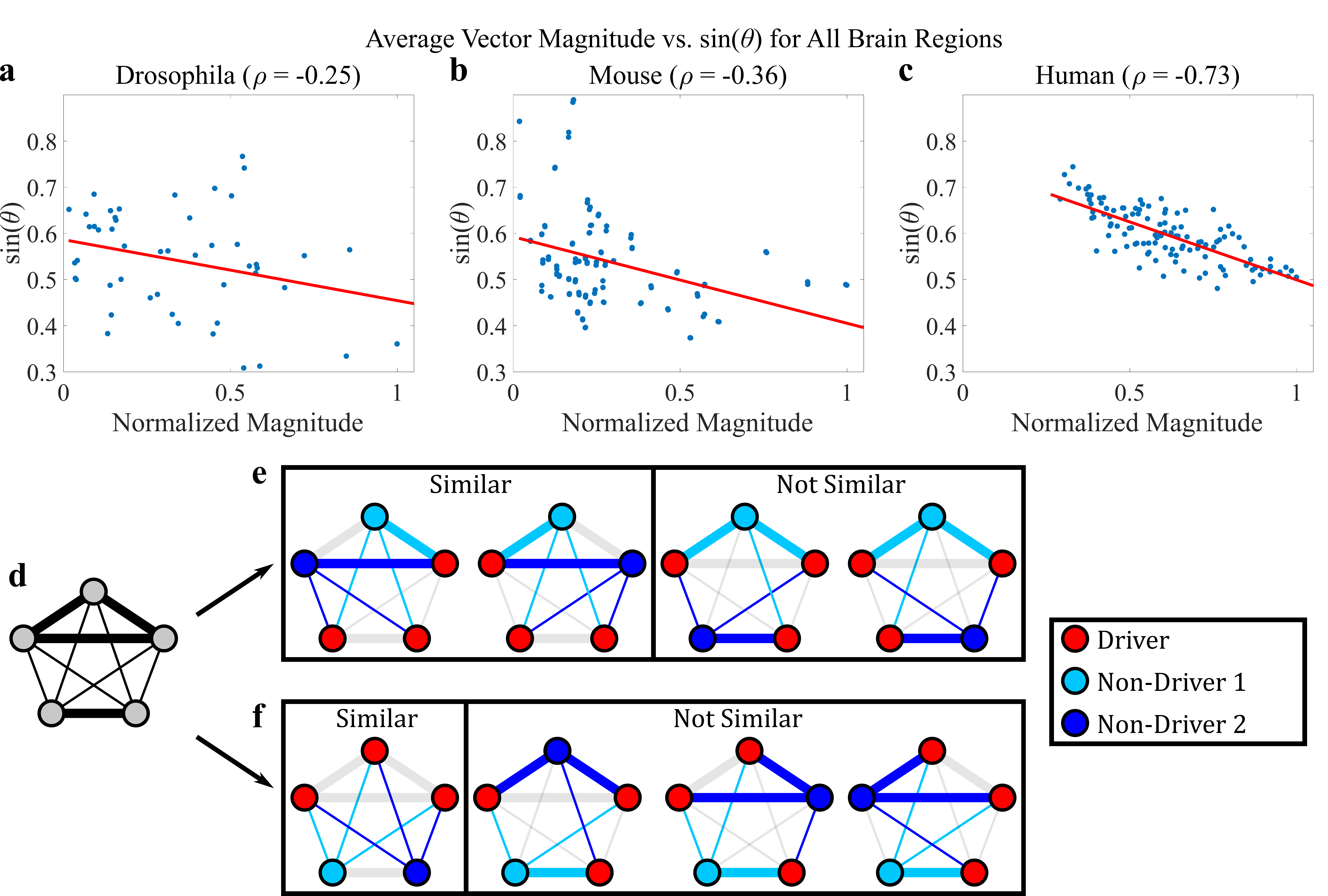}
	\caption{\textbf{Energetically Favorable Organization of Topological Features in Networks.} (\textbf{a}) Average $\sin(\theta_k)$ vs normalized $\|\bm{a}_k\|$ for each brain region across 10,000 random non-driver selections for a non-driver fraction of 0.2, along with best fit line (red) and corresponding Spearman correlation coefficient in the drosophila ($\rho = -0.25$, $p = 0.0748$), (\textbf{b}) mouse ($\rho = -0.36$, $p = 0.000125$), and (\textbf{c}) human ($\rho = -0.73$, $p \approx 0$). (\textbf{d}) Example toy network of 5 nodes displaying negative relationship between $\sin(\theta_k)$ vs. $\|\bm{a}_k\|$, with three strongly interconnected nodes at the top, and two strongly interconnected nodes at the bottom. (\textbf{e}) Representation of similarity in driver $\to$ non-driver connections between Non-Driver 1 (light blue, member of three strongly connected nodes) and all possible selections of Non-Driver 2 (blue). Across all 4 configurations, Non-Driver 1 has an average of 1.5 strong connections, and 2/4 similarly connected (small angle) configurations. (\textbf{f}) Similarity in driver $\to$ non-driver connections between Non-Driver 1 (light blue, member of two strongly connected nodes) and all selections of Non-Driver 2 (blue). Across all 4 configurations, Non-Driver 1 has an average of .75 strong connections, and 1/4 similarly connected configurations}
	\label{fig:mag_ang}
\end{figure}

To explore the relationship between $\|\bm{a}_k\|$ and $\sin(\theta_k)$ in brain networks, we selected 10,000 random permutations of non-drivers in the drosophila, mouse, and 10 human connectomes, at a non-driver fraction of 0.2. For each permutation, we calculated $\|\bm{a}_k\|$ and $\sin(\theta_k)$ for every non-driver. Then, we averaged $\|\bm{a}_k\|$ and $\sin(\theta_k)$ for each non-driver across all permutations, giving us an averaged magnitude $\|\bm{a}_k\|$ and $\sin(\theta_k)$ for each brain region in the drosophila, mouse, and each of 10 humans. Finally, we plotted the averaged $\sin(\theta_k)$ vs $\|\bm{a}_k\|$ for all brain regions in the drosophila, mouse, and the average across all 10 human subjects (Fig.~\ref{fig:mag_ang}a--c). We find little relationship between the averaged $\|\bm{a}_k\|$ and $\sin(\theta_k)$ in the drosophila (Spearman $\rho = -0.25$, $p = 0.0748$), a moderate negative relationship in the mouse ($\rho = -0.36$, $p = 0.000125$), and a strong negative relationship in the human ($\rho = -0.73$, $p \approx 0$). 

To graphically demonstrate how this negative $\sin(\theta_k)$ vs. $\|\bm{a}_k\|$ relation might arise in networks, we show a simple 5 node network with two communities of 3 and 2 strongly interconnected sets of nodes (Fig.~\ref{fig:mag_ang}d). We first look at the average $\|\bm{a}_k\|$ and $\sin(\theta_k)$ between a specific node in the 3 strongly interconnected set (Non-Driver 1, colored light blue in Fig.~\ref{fig:mag_ang}e) and all four permutations of one other non-driver (designated Non-Driver 2) as shown in Fig.~\ref{fig:mag_ang}e. We see that, because Non-Driver 1 is a member of 3 strongly interconnected nodes, it has on average 1.5 strong driver $\to$ non-driver edges across the 4 permutations, and 2/4 configurations where it and Non-Driver 2 are similarly connected (small angle). In contrast, we look at a specific node in the 2 strongly interconnected set (Non-Driver 1, colored light blue in Fig.~\ref{fig:mag_ang}f), and all four permutations of one other non-driver (Non-Driver 2, Fig.~\ref{fig:mag_ang}f). Here, we find that Non-Driver 1 only has an average of 0.75 strong driver $\to$ non-driver connections, and only 1/4 similarly connected configurations. Hence, on average, a non-driver among 3 strongly interconnected nodes will have stronger driver $\to$ non-driver connections (larger $\|\bm{a}_k\|$) and greater number of similarly connected configurations (smaller $\sin(\theta_k)$) than a non-driver among 2 strongly interconnected nodes.

\subsection{Network manipulation to facilitate control}

In the previous section, we segregated the energy contribution of network topology into two key components (magnitude and angle), and found that brain networks of more complex species were organized in more energetically favorable ways.  Here, we look to extend this concept to network modifications that lead to lower control energies. We focus on the effects of edge deletion since it is often useful in the study of biological systems such as brain \cite{Alstott2009}, metabolic \cite{Aristidou1994,Patnaik1994}, and gene regulatory \cite{Sander2014} networks. Perhaps counter-intuitively, we show that the deletion of certain edges improves the general controllability of the network. 

Ultimately, we seek to understand if, and to what degree, edge lesioning may be used therapeutically to improve a network's control profile. To that end, we quantified the effect of modifying each edge weight on the determinant, deleted edges that maximally increased the determinant, and demonstrated the corresponding impact on the total control energy. First, we let
$Q=\subscr{A}{21}\subscr{A}{21}^T$ be as in Eq.~\eqref{eq: energy determinant} and derived (Lemma~\ref{lemma: gram derivative L}) that
\begin{align}\label{eq: derivative}
  \frac{\partial}{\partial \subscr{A}{21}} \det(Q) = 2\det(Q)(Q^{-1} \subscr{A}{21}),
\end{align}
which characterized the rate of change of the determinant of $Q$ with respect to a change in any particular edge weight in $\subscr{A}{21}$. This metric was of particular utility in assessing the sensitivity of the control energy with respect to changes of the network's edge weights.

\begin{figure}[h!]
    \centering
    \includegraphics[width=1.0\columnwidth]{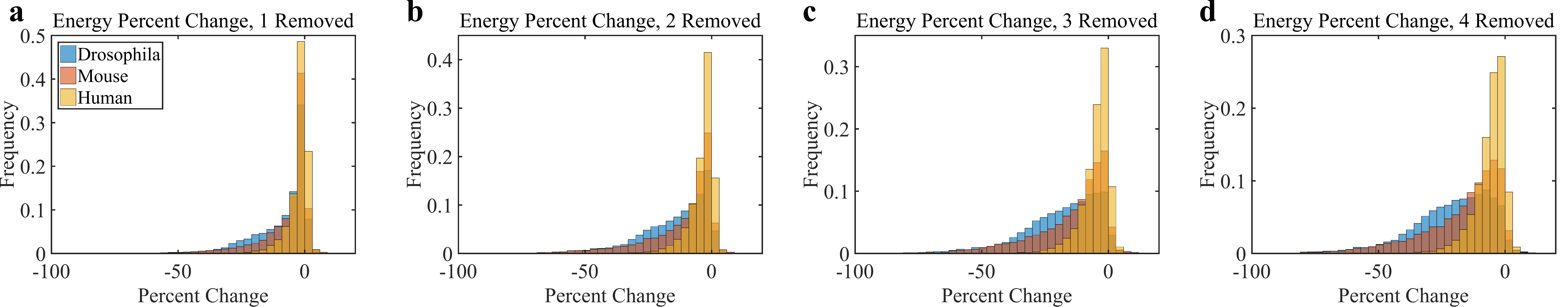}
    \caption{\textbf{Modifying the Drosophila, Mouse, and Human Connectomes to Decrease the Minimum Energy Required for Control.} (\textbf{a}) Distribution of percent change in control energy before and after deleting edges that maximally increase the determinant based on Eq.~\eqref{eq: derivative} over 10,000 control tasks, with initial states $\subscr{\bm{x}}{nd}(0) = 0$, $\subscr{\bm{x}}{d}(0) = 0$, and random final states $\subscr{\bm{x}}{nd}^* \in (-1,1)^M,\subscr{\bm{x}}{d}^* \in (-1,1)^N$. Non-drivers were randomly selected for a non-driver fraction of 0.2  in the drosophila, mouse, and human connectomes for 1 deletion, (\textbf{b}) 2 deletions, (\textbf{c}) 3 deletions, and (\textbf{d}) 4 deletions.}
    \label{fig:weights_remove}
\end{figure}

Then, we used Eq.~\eqref{eq: derivative} to make informed modifications to the network topology to increase or decrease the network control energy. First, we randomly selected 10,000 permutations of non-drivers at a non-driver fraction of 0.2, and designated the remaining regions as drivers. For each permutation, we extracted the block matrix $\subscr{A}{21}$, calculated $2\det(Q)(Q^{-1})\subscr{A}{21}$, and found the element $a_{ij} \neq 0$ yielding the largest change based on Eq.~\eqref{eq: derivative}. We then simulated an edge deletion by setting $a_{ij} = 0$, and we repeated the process to obtain networks of 1, 2, 3, and 4 deleted edges. Finally, for each permutation, we performed a random control task on the non-simplified network with initial states $\subscr{\bm{x}}{nd}(0) = 0$, $ \subscr{\bm{x}}{d}(0) = 0$, and final states  $\subscr{\bm{x}}{nd}^* \in (-1,1)^M,\subscr{\bm{x}}{d}^* \in (-1,1)^N$, and calculated the percent change in minimum energy between the pre- and post-modified networks for drosophila, mouse, and human connectomes Fig.~\ref{fig:weights_remove}a--d. As can be seen in Fig.~\ref{fig:weights_remove}a, the removal of a single weight can sometimes lead to more than a 50\% reduction in control energy, while the removal of four edges (Fig.~\ref{fig:weights_remove}d) can sometimes lead to more than an 80\% reduction in control energy.  We also note that the human connectome, which we showed was already energetically favorable wired, also had the smallest percent decrease in energy from edge-deletion.

Here we use the scaling property of the determinant to make small, strategic topological changes that drastically reduce the control energy. We show that the deletion of one edge can yield as much as a 50\% reduction in energy. We also show that, for the same topological modification, the drosophila experienced greater energy reduction than the mouse, which also experienced greater energy reduction than the human. This corresponds to the previous finding where, because brain networks of increasingly complex species are already energetically favorably wired, they may not experience as much improvement after modification. We emphasize that this was a purely topologically-motivated modification that did not cater to the specific end-state of the control task. This analysis may also yield a measure of how fragile or robust a specific network is to topological disruption, and introduces the perspective of targeted improvements in network controllability through topological changes.

\section{Discussion}

The control of networked systems is a critical frontier in science, mathematics, and engineering, as it requires a fundamental understanding of the mechanisms that drive network dynamics and subsequently offers the knowledge necessary to intervene in real-world systems to enhance or better their outcomes \cite{motter2015network}. While some theoretical predictions are beginning to be made in nonlinear network systems \cite{Cornelius2013}, the overwhelming majority of recent advances have been made in the context of linear control \cite{liu2011controllability,campbell2015topological,ruths2014control}. Yet, despite the significant efforts, some very basic intuitions regarding how edge weights impact control -- either locally around a certain node or globally throughout the whole system -- have remained elusive.  Here, we sought to address this gap by segregating network nodes into either drivers or non-drivers, and examining the dynamic interactions between them. We show that for a wide range of parameters, the minimum energy required to control a network is mostly a function of these directed bipartite connections, offering a fundamental theory of driver$\to$non-driver interactions. We apply this framework to the inter-areal connectomes of the mouse \cite{oh2014mesoscale,rubinov2015wiring}, drosophila \cite{shih2015connectomics}, and human to demonstrate that the predictions of the theory derived from the bipartite subgraph hold for the fully weighted, directed network. The work thus offers important insights into network analysis and design by delineating key topological principles in under-actuated network systems.

More specifically, we presented an equivalent first-order energy expression segregating the control energy into task-based and topology-based terms. We built upon our finding that similarly connected non-driver nodes decrease the gram determinant and increase the control energy by quantifying the connection similarity between non-driver $i$ and non-drivers $j \neq i$ in this topology term. Then we selected non-drivers with the topological goal of minimizing or maximizing this similarity, and showed that on average, networks with non-drivers that minimized this similarity required 2.5 -- 4 orders of magnitude less energy than those that maximized this similarity. We also showed that the connectomes of increasingly complex species were topologically organized to be energetically favorable. Ultimately, we have shown that there is an inherent topological contribution to the total control energy, which can be used to inform and analyze the selection of drivers and non-drivers. We concluded by using these principles to show that the deletion of one edge in many thousands of potential edges can yield as much as a 50\% reduction in energy, providing insights into how to target improvements in network controllability through fine-scale changes in topology.

\subsection{A novel conceptualization of network control}

A distinct advantage of our approach is the focus on a physically meaningful topological understanding of the principles governing network control. Although spectral analysis of a network's controllability Gramian \cite{TK:80} yields theoretically useful information about the overall behavior of the network under control \cite{pasqualetti2014controllability, yan2015spectrum}, it is not obvious how specific patterns of connectivity or selections of driver and non-driver nodes contribute to this behavior. Understanding this relationship is crucial when analyzing empirical biological networks such as the brain compared to purely mathematical networks, because the nodes and edges of a brain network often have known functions and perform known computations \cite{Lanteaume2007, Burgess2002}, and we are interested in understanding how these functions and computations modulate or influence one other. 

We address this gap in understanding between the control behavior and topology of networks through a simplified network only involving connections from driver to non-driver nodes. This simplification is quite generally motivated by recent work demonstrating that relatively sparse network representations of complex biological systems \cite{olhede2014network,park2015sparse,liu2016sparse} can contain much of the information needed to understand the system's structure and dynamics \cite{clauset2008hierarchical,yang2016predicting,pan2016predicting,zhu2015information,lu2015toward,navlakha2012network}.
More specifically, the simplification hard-codes the fact that energy can be transmitted directly from drivers to non-drivers along walks of length unity. By simplifying the complex network structure into driver $\to$ non-driver interactions, we reached a powerful closed-form bilinear mathematical approximation relating a network's topology to the total control energy. We found that for a range of matrix scales and fractions of non-driver nodes, this simplification well approximates the minimum control energy of the unsimplified network. This implies that control dynamics along the first-order connections from driver $\to$ non-driver nodes dominate the dynamics along other connections within the viable range of parameters. These results inform our understanding of how much first-order connections contribute to the overall dynamics of generic network control systems \cite{goni2014resting}.  

We used this approach to demonstrate that the similarity and strength of connections from driver to non-driver nodes are key topological principles that govern controllability. To reach this conclusion, we posited a geometric perspective of network control by showing that the control energy was inversely proportional to the determinant of the gram matrix $Q = \subscr{A}{21}\subscr{A}{21}^T$, and that this determinant was equal to the squared volume of the parallelotope formed by the vector of connections $\bm{a}_k$ into the $k$-th non-driver node. This principle allows us to make hypotheses about the function of a network given its structure. Brain regions that have very similar connection distributions form a parallelotope with a very small volume, implying that most state changes will be prohibitively energetically costly. Hence, the subspace of energetically viable state changes will be of fairly low dimension. However, very differentially connected regions will form a parallelotope with much larger volume, allowing the subspace of energetically viable state changes to be of higher dimension. These results inform the development of analytical constraints on the accessible state space of a networked system \cite{Cornelius2013}, particularly informing the set of states within which one might seek to push the brain using stimulation paradigms common in the treatment of neurological disorders and psychiatric disease \cite{chen2014harnessing,chrysikou2011noninvasive}. While many initial studies have examined unconstrained state spaces \cite{Gu2015,Betzel2016,muldoon2016stimulation}, understanding viable states and state trajectories is critical for the translation of these ideas into the clinic \cite{bassett2017emerging}.

Indeed, to provide further insights into the potential utility of these approaches in informing interventions in brain systems, we formally quantified the contribution of the network topology to the control energy as a sum of contributions from each non-driver, where each contribution was a function of the magnitude and angle of $\bm{a}_k$. This formulation allowed us to identify brain regions that were inherently costly to control, and therefore potentially should be chosen as stimulation targets with careful consideration. Moreover, we were able to use a greedy algorithm to find the set of $M$ non-drivers that maximized or minimized this topology-based energy component. Through this topology-based energy contribution, we obtained approximate answers to questions such as "which sets of brain regions are easiest \emph{versus} most difficult to control," and "which non-drivers should be designated drivers to most improve controllability." Importantly, these insights lay the groundwork for the optimization of stimulation sites in neural systems, a problem that has received very little theoretical treatment, and is considered one of the current critical challenges in neuroengineering \cite{johnson2013neuromodulation}.

Finally, we used the scaling principle of the gram determinant to make strategic, task-agnostic edge deletions that maximally increased the determinant. We saw that even the deletion of 1 edge occasionally produced a 50\% reduction in total control energy, while the deletion of 4 edges occasionally produced an 80\% reduction in total control energy. The first insight was that, even in an overdetermined, unsimplified system ($N>M$), a single edge deletion could produce such a profound improvement in the general controllability of a network. This sensitivity suggests that dynamical networks such as the brain can produce fairly drastic changes in dynamical behavior given minute changes in physiological topology, consistent with observations of critical dynamics in human and animal neurophysiology \cite{bassett2006adaptive,rubinov2011neurobiologically,shew2015adaptation,deco2012ongoing}. Moreover, these results also suggest that minor, targeted structural changes through concussive injury can lead to drastic changes in overall brain function \cite{hart2016graph,caegenberghs2016mapping,horn2016altered}, via altering the controllability landscape of the brain \cite{gu2016optimal}. The second interesting insight was that these topological modifications were task-agnostic edge deletions, signifying that even in a linear regime, the presence of an unfavorable edge can have a profoundly negative impact on the controllability of a network. We note that it is trivial to perform a similar analysis that takes into account the specific tasks $\bm{v}_1, \bm{v}_2$ by taking the derivative of the full energy term $E_{total}$ with respect to $\subscr{A}{21}$, which would optimize the network topology for a specific task, as studied in more detail in \cite{Betzel2016}.
 
\subsection{Cross-species comparison of controllability in structural brain networks}

Emerging neurotechnologies are uncovering the richness of brain connectivity with unprecedented detail \cite{bassett2016network}. The fine-scale maps produced by these efforts make quantiative cross-species comparison tractable in a way that was not possible in previous years \cite{heuvel2016comparative,reid2016cross}. Here, we take advantage of these advances to address the question of whether and how species differ in brain network controllability. Importantly, the more general question of whether brain network architecture in different species harbors similar or distinct topological attributes is not a new one \cite{bassett2006small,bassett2016small,bassett2010efficient,kaiser2006nonoptimal}. The majority of work has focused on reporting cross-species similarities, while few have addressed the question of evolutionary drivers explaining differences across species capable of more or less complex function \cite{herculano2012remarkable}. Here we find a monotonic gradation in the mapping between edge strength and topological angle across the 3 species, suggesting that increasingly complex species are more energetically favorably wired. Interestingly, the human, in addition to being most energetically favorably wired, also had the smallest percent decrease in energy following edge-deletion. These results point to an advantage of the human brain in supporting diverse network dynamics with small energetic costs, while remaining unexpectedly robust to perturbations. It will be interesting in the future to expand this analysis to the connectomes of other species as they become available, and also to examine brain network robustness in individuals sustaining traumatic brain injury \cite{gu2016optimal}.

\subsection{Utility in informing control of brain networks}

In this paper, we apply our theoretical framework to real-world data collected via tract-tracing in drosophila and mouse, and via diffusion imaging in human. Such brain networks represent particularly important contexts in which to understand control \cite{bassett2016network,bassett2017emerging}. Even at the microscale of individual neurons, neural control engineering \cite{schiff2011neural} seeks to identify minimal energy control \cite{nabi2013minimum}, that can potentially be used to explain homeostatic mechanisms controlling abonormal bursts of activity \cite{wiles2016autaptic} or to develop exogeneous control strategies to terminate bursts \cite{wilson2014hamilton}. Open questions surround the role of symmetries \cite{whalen2012observability,whalen2015observability} or synchronizability \cite{tang2016structural} within the network that may constrain these control properties. At the macroscale of centimeter-sized brain regions, network control offers a novel perspective on how the brain controls itself, a characteristic known to cognitive neuroscientists as \emph{cognitive control} \cite{Gu2015}, as well as which mental states might be preferred \cite{Betzel2016}, and how both features may be altered following traumatic brain injury \cite{gu2016optimal}. Indeed, in clinical populations the need for realistic, low-energy control systems is particularly pressing, for example to restore neural function in Parkinson's disease \cite{santaniello2015therapeautic}, to supress bursting activity in coma \cite{ching2012neurophysiological,ching2013real}, and to control the distributed propagation of seizures \cite{ching2012distributed} in epilepsy, which has become known as an inherently network-based disorder \cite{burns2014network,khambhati2015dynamic,khambhati2016virtual}. Our results offer a novel framework in which to address these questions and challenges, as well as tools to design interventions (the strengthening and weakening of neural connections) to facilitate optimal control and enhance therapeutic benefit \cite{bassett2017emerging}.

\subsection{Theoretical considerations and methodological limitations}

The work is built on several important assumptions that bring with them significant theoretical considerations. First, we only consider first-order connections from driver nodes to non-driver nodes. In other words, we study paths of length $1$ between drivers and non-drivers; we do not study the propagation of control energy through longer paths or walks in the network. This is an inherent limitation of the work, as it is clear from prior studies that the strength of long walks through the network has a non-trivial impact on the control energy \cite{Betzel2016,gu2016optimal}. To address this limitation, we demonstrate that this simplification enables us to better understand the dependence of control energy on network topology. Moreover, we show that these first-order control dynamics can offer reasonable approximations for non-simplified networks constructed from real-world neuroimaging data in drosophila, mouse, and human. 

Second, we assume that these networks adhere to linear dynamics \cite{liu2011controllability, muller2011few, yan2015spectrum}, which may limit the applicability of the results to (i) linear systems, or (ii) nonlinear systems for which control is sought in short time intervals, enabling a linearization of the dynamics around the operating point \cite{luenberger1979introduction}.  Moreover, in the context of our application to understand neural architecture, this choice is consistent with prior brain network control studies, such as \cite{Gu2015,Betzel2016,muldoon2016stimulation}, as well as prior mathematical modeling studies on human neuroimaging data \cite{galan2008network,honey2009predicting}.  

Third, we assume that the input functions are chosen in such a way as to minimize both the control energy and the distance of the current state from the target state \cite{Betzel2016,gu2016optimal}. Importantly, these assumptions represent the best case scenario, offering a lower-bound on the relationships between topology and control energy. Additional variables of interest to examine in future studies include the time scales of control and the tortuosity of the trajectory.

Finally, as with all empirical data, the connectomes of the drosophila, mouse, and human are fundamentally estimates of the true large-scale brain connectivity, and emerging neurotechnologies will likely offer increasingly accurate estimates. A specific limitation of the human connectome that is important to mention is that the diffusion imaging data must be submitted to sophisticated tractography algorithms to construct region-to-region estimates of structural connectivity. While evolving at a swift pace, these algorithms may still report spurious tracts or fail to report existing tracts \citep{thomas2014anatomical, reveley2015superficial,pestilli2014evaluation}. However, these issues are somewhat mitigated by the fact that we use an exceptionally high-resolution scan, capitalizing on a multiband sequence taking place over 55 minutes, and estimating diffusion over 705 directions, thereby increasing the resolution of the data by an order of magnitude over most existing studies (which estimate diffusion over 30-64 diffusion directions).

\subsection{Conclusion and future directions}

In closing, we note that the natural direction in which to take this work will be to include higher order interactions in the bipartite framework, and further to expand the bipartite framework to include driver$\to$driver and non-driver$\to$non-driver interactions. Moreover, it would be interesting to apply this reduced framework to random graphs and other well-known benchmarks -- both from a mathematical perspective \cite{bollobas1985random} and also in the context of neural systems \cite{klimm2014resolving,sizemore2016classification} -- to better understand the phenotypes present in those graph ensembles. Third and finally, informing the design of new networks with these tools may be particularly useful in neuromorphic computing \cite{pedroni2016mapping,pfeil2013six}, material science \cite{papadopoulos2016evolution,giusti2016topological}, and other contexts where optimal control of physical systems is of paramount importance.

\subsection{Acknowledgements} JK acknowledges support from NIH T32-EB020087. JMS and DSB acknowledge support from the John D. and Catherine T. MacArthur Foundation, the Alfred P. Sloan Foundation, the U.S. Army Research Laboratory and the U.S. Army Research Office through contract numbers W911NF-10-2-0022 and W911NF-14-1-0679, the National Institute of Health (2-R01-DC-009209-11, 1R01HD086888-01, R01-MH107235, R01-MH107703, R01MH109520, 1R01NS099348 R21-M MH-106799, and T32-EB020087), the Office of Naval Research, and the National Science Foundation (BCS-1441502, CAREER PHY-1554488, BCS-1631550, and CNS-1626008). AEK and JMV acknowledge support from the U.S. Army Research Laboratory contract number W911NF-10-2-0022. FP acknowledges support from the National Science Foundation (BCS-1430280 and BCS 1631112). The content is solely the responsibility of the authors and does not necessarily represent the official views of any of the funding agencies.

\newpage
\section{Supplement}

\subsection{Connectome Data}

\emph{Drosophila Connectome.} The full reconstruction of the Drosophila connectome can be found in the FlyCircuit 1.1 database  \cite{chiang2011three,shih2015connectomics}. This database contains images of 12,995 neurons, as well as their projections, that are characteristic of the Drosophila female. In this database, each neuron was labeled using green fluorescent protein (GFP) and its location was estimated from 3-dimensional images that were co-registered to a template using a rigid linear transform. To obtain a mesoscale representation of this fine-scale data, neurons were assigned to one of 49 local populations, based on their morphology and known functions. Following the original work from Shih and colleagues, we treated each of these 49 populations as the nodes of the network, and we treated the directed, weighted edges between populations as network edges \cite{shih2015connectomics}. 

\emph{Mouse Inter-Areal Connectome Data.} In addition to drosophila connectome, we also analyzed the inter-areal connectome of the mouse. In particular, we use the exact network studied in \cite{rubinov2015wiring}, which was reconstructed from original tract-tracing data recently released by the Allen Brain Institute \cite{oh2014mesoscale}. The entire brain was separated into 112 regions, which we treat as network nodes. Each pair of regions was then linked by directed edges that encoded the presence or absence of inter-regional projections. The weight of each edge was defined by the number of projections normalized by the volumes of the two regions being connected. 

\emph{Human Diffusion Imaging Data.} Ten healthy adult human subjects (m) were imaged as part of an ongoing data collection effort at the University of Pennsylvania; the subjects provided informed consent in writing, in accordance with the Institutional Review Board of the University of Pennsylvania. All scans were acquired on a Siemens Magnetom Prisma 3 Tesla scanner with a 64-channel head/neck array at the University of Pennsylvania. Each data acquisition session included both a diffusion spectrum imaging (DSI) scan as well as a high-resolution T1-weighted anatomical scan. The diffusion scan was 730-directional with a maximum $b$-value of 5010s/mm$^{2}$ and TE/TR = 102/4300 ms, which included 21 $b=0$ images. Matrix size was 144$\times$144 with a slice number of 87. Field of view was 260$\times$260mm$^2$ and slice thickness was 1.80mm. Acquisition time per DTI scan was 53:24min, using a multi-band acceleration factor of 3. The anatomical scan was a high-resolution three-dimensional T1-weighted sagittal whole-brain image using a magnetization prepared rapid acquisition gradient-echo (MPRAGE) sequence. It was acquired with TR = 2500 ms; TE=2.18 ms; flip angle = 7 degrees; 208 slices; 0.9mm thickness.

DWI is highly sensitive to subject movement\citep{Yendiki:2013ez}, which can cause significant distortions in the reconstructed ODFs if not corrected. Motion correction is typically applied by determining an affine or non-linear transform to align each DWI volume to a reference derived from the high-signal $b=0$ images. The high b-values used in DSI present a problem for this approach, as the low signal in many of the volumes leads to poor registration. To address this, we interspersed $b=0$ volumes in the scan sequence, one for every 35 volumes. An initial average template was produced by averaging the $b=0$ images together and then improved by registering the $b=0$ images to the initial template and re-averaging. Each $b=0$ was finally re-registered to the improved template, and then each volume in the DSI scan was then motion corrected by applying the transformation calculated for the closest $b=0$ volume. Motion correction also impacts the effective $b$-matrix directions since the rotated images are no longer aligned with the scanner; therefore the transforms applied to motion correct each volume were also used to rotate the corresponding $b$-vectors.\citep{Leemans:2009jn} The processing pipeline was implemented using Nipype\citep{Gorgolewski:2011ex} with registration performed using the Advanced Normalization Tools (ANTs)\citep{Avants:2011kk}.

Using DSI-Studio (http://dsi-studio.labsolver.org), orientation density functions (ODFs) within each voxel were reconstructed from the corrected scans using GQI \citep{Yeh:2010jq}. We then used the reconstructed ODFs to perform a whole-brain deterministic tractography using the derived QA values in DSI-Studio \citep{Yeh:2013fa}. We generated 1,000,000 streamlines per subject, with a maximum turning angle of 35 degrees\citep{Bassett2011structure} and a maximum length of 500mm\citep{Cieslak2014}. By holding the number of streamlines between participants constant, we use the number of streamlines that connect brain region pairs as an estimate of the strength of the connection and examine individual variability in structural connectivity\citep{Griffa:2013gn}.

To examine the relationship between structural connectivity and individual differences in learning rate, we constructed networks for each subject where nodes are atlas regions and edges are the measured connection strength between region pairs \citep{Hagmann:2008gd}. The nodes of the network were derived from spatially-defined regions of a brain atlas. We chose the anatomically-defined AAL atlas, originally developed in Statistical Parametric Mapping (SPM) \citep{TzourioMazoyer:2002bi}, which divides each brain hemisphere into 45 regions. We used a version in MNI-space that was then warped into subject-specific space using ANTs. Edges of the network were derived from streamlines that started and ended between the region pair and excluded streamlines that passed through one or both of the regions.

\subsection{Mathematical Framework}
Here we reiterate our mathematical notation and assumptions in greater detail, and provide lemmas for the main results. Consider a network represented by the directed graph $\mc G = (\mc V, \mc E)$, where $\mc V = \until{n}$ and $\mc E \subseteq \mc V \times \mc V$ are
the sets of network vertices and edges, respectively. Let
$a_{ij} \in \real$ be the weight associated with the edge
$(i,j) \in \mc E$, and let $A = [a_{ij}]$ be the weighted adjacency
matrix of $\mc G$. We associate a real value (\emph{state}) with each
node, collect the nodes' states into a vector (\emph{network state}),
and define the map $\map{\bm{x}}{\real_{\ge 0}}{\real^n}$ to describe the
evolution (\emph{dynamics}) of the network state over time. We
let the network dynamics be linear and time invariant, as described by
the equation

\begin{align}\label{eq: autonomous dynamics L}
\dot{\bm{x}} = A \bm{x} .
\end{align}

We are particularly interested in characterizing how the network
structure $\mc G$ influences the control properties of the dynamical
system \eqref{eq: autonomous dynamics L}. To this aim, we assume that a
subset of $N$ nodes, called drivers, is independently manipulated by
external controls and, without loss of generality, we reorder the
network nodes such that the $N$ drivers come first. Thus,
the network dynamics with controlled drivers read as

\begin{align}\label{eq: control dynamics supp}
\begin{bmatrix}
\subscr{\dot{\bm{x}}}{d} \\ 
\subscr{\dot{\bm{x}}}{nd}
\end{bmatrix}
=
\begin{bmatrix}
A_{11} & A_{12}\\
A_{21} & A_{22}
\end{bmatrix}
\begin{bmatrix}
\subscr{\bm{x}}{d} \\ 
\subscr{\bm{x}}{nd}
\end{bmatrix}
+
\begin{bmatrix}
I_N \\ 0
\end{bmatrix}
\bm{u},
\end{align}

\noindent where $\subscr{\bm{x}}{d}$ and $\subscr{\bm{x}}{nd}$ are the state vectors of
the driver and non-driver nodes, $A_{11} \in \real^{N \times N}$,
$M = n - N$, $A_{12} \in \real^{N \times M}$,
$A_{21} \in \real^{M \times N}$, $A_{22} \in \real^{M \times M}$,
$I_N$ is the $N$-dimensional identity matrix, and
$\map{\bm{u}}{\real_{\ge 0}}{\real^N}$ is the control input.

We say that the network is \emph{controllable} at time
$T \in \real_{\ge 0}$ if, for any pair of states $\subscr{\bm{x}}{d}^*$ and
$\subscr{\bm{x}}{nd}^*$, there exists a control input $u$ for the
dynamics \eqref{eq: control
	dynamics L} such that $\subscr{\bm{x}}{d}(T) = \subscr{\bm{x}}{d}^*$ and
$\subscr{\bm{x}}{nd}(T) = \subscr{\bm{x}}{nd}^*$. We refer the interested reader
to \cite{TK:80} for a detailed discussion and rigorous conditions for
the controllability of a system with linear dynamics. Finally, we
define the energy of $u$ as 
\begin{align*}
\textup{E} (\bm{u}) = \sum_{i = 1}^N \underbrace{\int_0^T  u_i
	(t)^2 dt}_{\textup{E}_i} ,
\end{align*}
where $u_i$ is the $i$-th component of $\bm{u}$. In what follows we
characterize how the network topology and weights determine the
control energy needed for a given control task. We restrict our
analysis to a class of bipartite networks, as specified in the
following assumptions. We remark that our assumptions, although
restrictive, allow us to thoroughly predict how driver $\to$
non-driver connections facilitate or inhibit network control even in
more complex network models (see Section \ref{sec: validation}).

\begin{assumption}\label{assumption: initial states}
	The network initial state satisfies $\subscr{\bm{x}}{d}(0)=0$ and
	$\subscr{\bm{x}}{nd}(0) = 0$. \oprocend
\end{assumption}

\begin{assumption}\label{assumption: network edges}
	The network $\mc G$ contains only edges from the drivers to the
	non-drivers, that is, $A_{11} = 0$, $A_{22} = 0$, and $A_{12} =
	0$.
	Thus, the dynamics \eqref{eq: control dynamics L} simplify to
	$\subscr{\dot{\bm{x}}}{d} = \bm{u}(t)$, and
	$\subscr{\dot{\bm{x}}}{nd} = A_{21} \subscr{\bm{x}}{d}$.
	\oprocend
\end{assumption}

\begin{figure}
	\centering
	\includegraphics[width=.7\columnwidth]{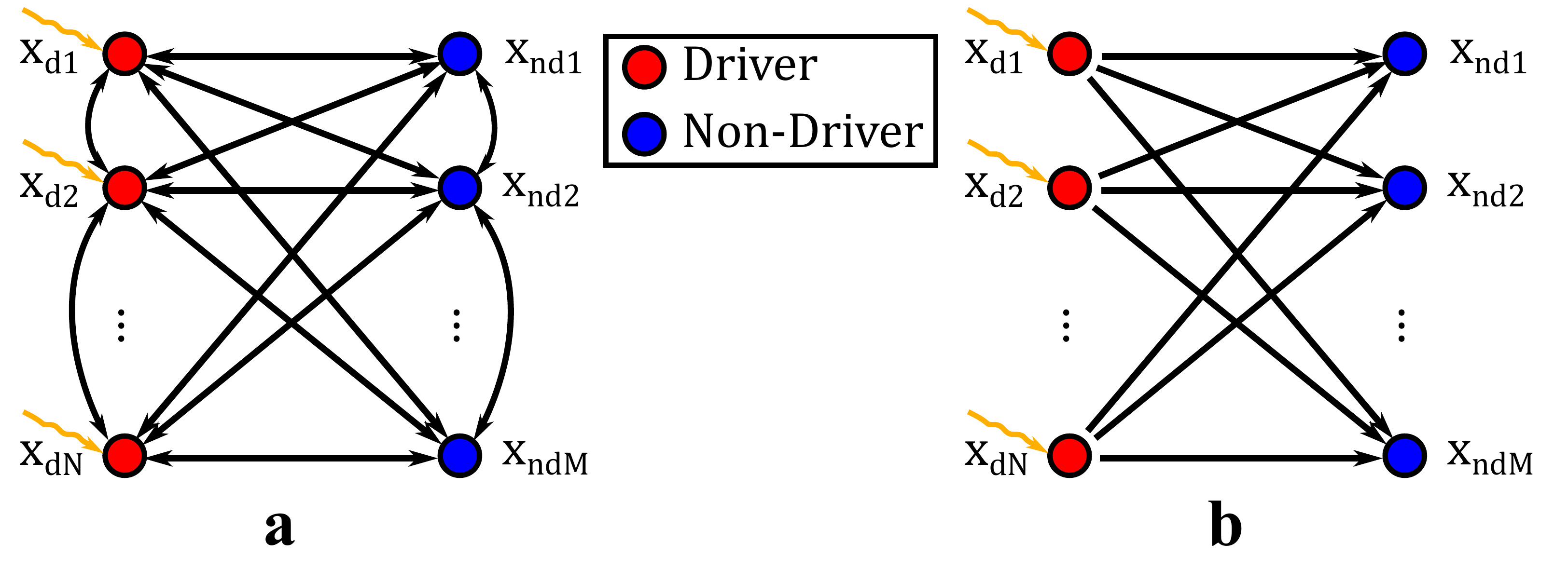}
	\caption{\textbf{Illustrations of Different Linear Network Control
			Frameworks}. (\textbf{a}) Network schematic of an unsimplified network with $N$ driver nodes ($\subscr{x}{d}$, in red), and $M$ non-driver nodes ($\subscr{x}{nd}$, in blue), with external control inputs ($u_i(t)$) represented as orange arrows.
		(\textbf{b}) Network schematic representing simplified network with only connections from drivers to non-drivers.}
	\label{fig: general_framework L}
\end{figure}

An example of the original network (Fig. \ref{fig: general_framework L}a) and the network satisfying our assumption (Fig. \ref{fig:
	general_framework L}b) are shown. From assumptions \ref{assumption: initial
	states} and \ref{assumption: network edges} we readily observe that

\begin{align}\label{eq: solution state L}
\begin{split}
\subscr{\bm{x}}{d} (t) &= \int_0^t \bm{u}(\tau) \,d\tau , \text{ and }\\
\subscr{\bm{x}}{nd} (t) &= A_{21} \int_0^t \subscr{\bm{x}}{d} (\tau) \,d\tau =
A_{21} \int_0^t \int_0^\tau \bm{u} (\tau_1) \,d\tau_1 \, d\tau .
\end{split}
\end{align}

\noindent We see that $\subscr{\bm{x}}{nd}$ provides an integral constraint to $\subscr{\bm{x}}{d}$, and represent the specific values of the constraints as

\begin{align}\label{eq: integral constraint L}
\int_0^t \subscr{\bm{x}}{d}(\tau) d\tau = \bm{C},
\end{align}

\noindent where $\bm{C} \in \real^{N \times 1}$ is an $N$-dimensional vector of real-valued constants. Furthermore, the set of controllable states can be characterized as
follows. For a matrix $M$, let $\Image (M)$ and $\text{Rank}(M)$
denote the image and rank of $M$, respectively \cite{CDM:01}.

\begin{lemma}{\bf \emph{(Controllability)}}\label{lemma: controllable
		subspace L}
	The network \eqref{eq: control dynamics L} is controllable if and only
	if $\text{Rank}(A_{21}) = M$. Furthermore, the set of controllable
	states is 
	$\Image \left(
	\begin{bmatrix}
	I_N & 0\\
	0 & A_{21}
	\end{bmatrix}
	\right)
	$.
\end{lemma}
\medskip
\begin{proof}
	Notice that the controllability matrix of \eqref{eq: control
		dynamics L} is
	\begin{align*}
	\mc C =
	\begin{bmatrix}
	B & AB
	\end{bmatrix}
	=
	\begin{bmatrix}
	I_N & 0\\
	0 & A_{21}
	\end{bmatrix}
	,
	\end{align*}
	and recall that a state is controllable if and only if it belongs to
	the range space of the controllability matrix.
\end{proof}

\begin{lemma}{\bf \emph{(Minimum Energy Control Input)}}\label{lemma: minimum control law L}
The $\supscr{i}{th}$ driver trajectory $\subscr{x}{d\textit{i}}(t)$ that minimizes the control energy takes the form $\subscr{x}{d\textit{i}}(t) = a_it^2+b_it$
\end{lemma}
\medskip
\begin{proof}
	Recall from \eqref{eq: control dynamics L} and assumption \ref{assumption: network edges} that $\subscr{\dot{x}}{d\textit{i}}(t) = u_i(t)$. We minimize the energy
	
	\begin{align*}
	E_i &= 
	\displaystyle \min_{u_i} \int_0^T u_i(t)^2 dt \\
	&= \displaystyle \min_{\subscr{x}{d\textit{i}}} \int_0^T \subscr{\dot{x}}{d\textit{i}}(t)^2 dt,
	\end{align*}
	
	\noindent where $\subscr{x}{d\textit{i}}(0) = 0, \subscr{x}{d\textit{i}}(T) = \subscr{x}{d\textit{i}}^*$. From \eqref{eq: integral constraint L}, $\subscr{x}{d\textit{i}}$ is also subject to some integral constraint
	
	\begin{align*}
	\int_0^T \subscr{x}{d\textit{i}}(t)dt &= C_i,
	\end{align*}
	
	\noindent where $C_i$ is the $i$th element of $\bm{C}$. We see this naturally takes the form of the isoperimetric problem in the calculus of variations, which finds
	
	\begin{align*}
	\displaystyle \min_{\subscr{x}{d\textit{i}}} \int_a^b F(t,\subscr{x}{d\textit{i}},\dot{x}\subscr{}{d\textit{i}}) dt,
	\end{align*}
	
	\noindent where $F(t,\subscr{x}{d\textit{i}},\dot{x}\subscr{}{d\textit{i}}) = \dot{x}\subscr{}{d\textit{i}}(t)^2$, $a = 0$, and $b=T=1$, constrained by
	
	\begin{align*}
	\int_a^b G(t,\subscr{x}{d\textit{i}},\dot{x}\subscr{}{d\textit{i}})dt = 0,
	\end{align*}
	
	\noindent where $G(t,\subscr{x}{d\textit{i}},\dot{x}\subscr{}{d\textit{i}}) = \subscr{x}{d\textit{i}}(t) - C_i$. The trajectory $\subscr{x}{d\textit{i}}^*(t)$ which locally minimizes the cost function must satisfy the necessary (Euler-Lagrange) and sufficient (Jacobi) conditions. The Euler-Lagrange equation reads
	
	\begin{align*}
	\frac{d}{dt}\frac{\partial}{\partial \dot{x}} (F + \lambda G) &= \frac{\partial}{\partial x}(F + \lambda G),
	\end{align*}
	
	\noindent which, after substituting $F$ and $G$, yields 
	
	\begin{align*}
	\subscr{\ddot{x}}{d\textit{i}}^*(t) &= \frac{\lambda}{2},
	\end{align*}
	
	\noindent to give the only extremal solution satisfying assumption \ref{assumption: initial states}
	
	\begin{align*}
	\subscr{x}{d\textit{i}}^*(t) &= \frac{\lambda}{2} t^2 + bt,
	\end{align*}
	
	\noindent where $\lambda$ is the lagrange multiplier. Because $(F + \lambda G)_{xx} = (F + \lambda G)_{x \dot{x}} = 0$, and $(F+\lambda G)_{\dot{x}\dot{x}} = 2$, the Jacobi condition becomes
	
	\begin{align*}
	\int_0^1 \dot{\eta}^2 (F+\lambda G)_{\dot{x}\dot{x}} dt = 2\int_0^1 \dot{\eta}^2 dt \geq 0,
	\end{align*}
	
	\noindent which holds true for any arbitrary smooth function $\eta$, where $\eta(0) = \eta(1) = 0$. As $\subscr{x}{d\textit{i}}^*(t) = at^2 + bt$ is the only extremal function, and is also minimum, $\subscr{x}{d\textit{i}}$ is the global minimum of the constrained control energy. 
	
\end{proof}

\begin{lemma}{\bf \emph{(Minimum Control Energy)}}\label{lemma: minimum control energy L}
	The required control energy for the $i^{th}$ driver is $E_i = 12C_i^2 - 12C_i\subscr{x}{d\textit{i}} + 4\subscr{x}{d\textit{i}}$.
\end{lemma}
\medskip
\begin{proof}
	We recall that the energy required to drive $\subscr{x}{d\textit{i}}$ is
	
	\begin{align*}
	E_i &= \int_0^{T=1} \dot{x}\subscr{}{d\textit{i}}^2(t) dt\\
	&= \int_0^1 4a_it^2 + 4a_ib_it + b_i^2 dt\\
	&= \frac{4}{3}a_i + 2a_ib_i + b_i^2.
	\end{align*}

	We solve for $a_i$ and $b_i$ via the final state and integral constraint to yields equations
	
	\begin{align*}
	\subscr{x}{d\textit{i}}(T=1) = a_i+b_i &= \subscr{x}{d\textit{i}}^*\\
	\int_0^{T=1}\subscr{x}{d\textit{i}}(t)dt = \frac{1}{3}a_i + \frac{1}{2}b_i &= C_i,
	\end{align*}
	
	\noindent from which we get
	
	\begin{align*}
	a_i &= 3\subscr{x}{d\textit{i}} - 6C_i\\
	b_i &= 6C_i - 2\subscr{x}{d\textit{i}}.
	\end{align*}
	
	substituting $a_i$ and $b_i$ into the equation for $E_i$, we get
	
	\begin{align*}
	E_i = 12C_i^2 - 12C_i\subscr{x}{d\textit{i}} + 4\subscr{x}{d\textit{i}}^2
	\end{align*}

\end{proof}

\begin{lemma}{\bf \emph{(Total Control Energy)}}\label{lemma: total control energy L}
	The total control energy is $E_{total} = 12 \bm{v}_1^T Q^{-1} \bm{v}_1 + \bm{v}_2^T\bm{v}_2$, where $\bm{v}_1 = \subscr{\bm{x}}{nd}^*-\frac{1}{2}\subscr{A}{21}\subscr{\bm{x}}{d}^*$, $\bm{v}_2 = \subscr{\bm{x}}{d}^{*T}\subscr{\bm{x}}{d}^*$, and $Q = \subscr{A}{21}\subscr{A}{21}^T$
\end{lemma}
\medskip
\begin{proof}
	
	Here, we use the method of Lagrange multipliers to minimize the total energy $f(\bm{C})$ as a function of $\bm{C}$ given in \eqref{eq: integral constraint L}, constrained by $g(\bm{C})$ given by \eqref{eq: solution state L}. We can write the total energy as
	
	\begin{align*}
	f(C) = E(\bm{u}) &= \sum_{i=1}^N E_i\\
	&= \sum_{i=1}^N 12C_i^2 - 12C_i\subscr{x}{d\textit{i}} + 4\subscr{x}{d\textit{i}}^2\\
	&= 12\bm{C}^T\bm{C} - 12\bm{C}^T\subscr{\bm{x}}{d} + 4\subscr{\bm{x}}{d}^T\subscr{\bm{x}}{d},
	\end{align*} 
	
	\noindent with $M$ constraining equations, the set of which are given by
	
	\begin{align*}
	g(\bm{C}) &= \subscr{\bm{x}}{nd}^* - \subscr{A}{21}\int_0^T\subscr{\bm{x}}{d}(\tau)d\tau\\&= \subscr{\bm{x}}{nd}^* - \subscr{A}{21} \bm{C}\\
	&= 0,
	\end{align*}
	
	\noindent where the $k^{th}$ constraint $g_k(\bm{C})$ is given by the $k^{th}$ row of $g(\bm{C})$. The method of Lagrange multipliers defines the Lagrangian given by
	
	\begin{align*}
	\mathcal{L}(\bm{C},\lambda_1,\dotsm,\lambda_M) &= f(\bm{C}) + \sum_{k=1}^M \lambda_k g_k(\bm{C})\\
	&= f(\bm{C}) + g(\bm{C})^T \bm{\lambda}\\
	&= 12\bm{C}^T\bm{C} - 12\bm{C}^T\subscr{\bm{x}}{d}^* -  \bm{C}^T\subscr{A}{21}^T\bm{\lambda} + \subscr{\bm{x}}{nd}^*\bm{\lambda} +  4\subscr{\bm{x}}{d}^{*T}\subscr{\bm{x}}{d}^*\\
	&= \bm{C}^T(12\bm{C} - 12\subscr{\bm{x}}{d}^* - \subscr{A}{21}^T \bm{\lambda}) +  \subscr{\bm{x}}{nd}^* \bm{\lambda} +  4\subscr{\bm{x}}{d}^{*T}\subscr{\bm{x}}{d}^*,
	\end{align*}
	
	\noindent where $\lambda_k$ is the $k^{th}$ Lagrange multiplier to compose $\bm{\lambda} \in \real^{M \times 1}$, and sets the gradient of the Lagrangian to 0
	
	\begin{align*}
	\nabla_{\bm{C}} \mathcal{L}(\bm{C},\lambda_1,\dotsm,\lambda_M) &= 24\bm{C} - 12\subscr{\bm{x}}{d} - \subscr{A}{21}^T \bm{\lambda}\\
	&= 0,
	\end{align*}
	
	\noindent which allows us to solve for $\bm{C}$ with respect to $\bm{\lambda}$
	
	\begin{align*}
	\bm{C} = \frac{1}{24}\subscr{A}{21}^T \bm{\lambda} + \frac{1}{2}\subscr{\bm{x}}{d}^*
	\end{align*}
	
	By substituting $\bm{C}$ into the total energy equation and grouping terms, we get a preliminary formulation of $E(\bm{u})$ with respect to $\bm{\lambda}$
	
	\begin{align*}
	E(\bm{u}) &= \frac{12}{24^2} \bm{\lambda}^T \subscr{A}{21} \subscr{A}{21}^T \bm{\lambda} + \left(\frac{12}{24} - \frac{12}{24}\right) \subscr{\bm{x}}{d}^{*T}\subscr{A}{21}^T \bm{\lambda} + \left(3-6+4\right) \subscr{\bm{x}}{d}^{*T}\subscr{\bm{x}}{d}^*\\
	&= \frac{\bm{\lambda}^T \subscr{A}{21}\subscr{A}{21}^T \bm{\lambda}}{48} + \subscr{\bm{x}}{d}^{*T}\subscr{\bm{x}}{d}^*.
	\end{align*}
	
	To solve for $\bm{\lambda}$, we substitute the expression for $\bm{C}$ into our constraint equations $g(\bm{C})$ to yield
	
	\begin{align*}
	g(\bm{C}) &= \subscr{\bm{x}}{nd}^* - \subscr{A}{21} \bm{C}\\
	&= \subscr{\bm{x}}{nd}^* - \frac{1}{24} \subscr{A}{21}\subscr{A}{21}^T \bm{\lambda} - \frac{1}{2} \subscr{A}{21}\subscr{\bm{x}}{d}^*\\
	&= 0,
	\end{align*}
	
	\noindent and solve for $\bm{\lambda}$
	
	\begin{align*}
	\bm{\lambda} = 24(\subscr{A}{21}\subscr{A}{21}^T)^{-1} (\subscr{\bm{x}}{nd}^* - \frac{1}{2}\subscr{A}{21}\subscr{\bm{x}}{d}^*).
	\end{align*}
	
	\noindent Substituting $\bm{\lambda}$ into $E(\bm{u})$, we get
	
	\begin{align*}
	E(\bm{u}) &= 12(\subscr{\bm{x}}{nd}^* - \frac{1}{2}\subscr{A}{21}\subscr{\bm{x}}{d}^*)^T(\subscr{A}{21}\subscr{A}{21}^T)^{-1}(\subscr{A}{21}\subscr{A}{21}^T)(\subscr{A}{21}\subscr{A}{21}^T)^{-1}(\subscr{\bm{x}}{nd}^* - \frac{1}{2}\subscr{A}{21}\subscr{\bm{x}}{d}^*) + \subscr{\bm{x}}{d}^{*T}\subscr{\bm{x}}{d}^*\\
	&= 12(\subscr{\bm{x}}{nd}^* - \frac{1}{2}\subscr{A}{21}\subscr{\bm{x}}{d}^*)^T(\subscr{A}{21}\subscr{A}{21}^T)^{-1}(\subscr{\bm{x}}{nd}^* - \frac{1}{2}\subscr{A}{21}\subscr{\bm{x}}{d}^*) + \subscr{\bm{x}}{d}^{*T}\subscr{\bm{x}}{d}^*\\
	\end{align*}
	
\end{proof}

\begin{lemma}{\bf \emph{(Derivative of Gram Matrix)}}\label{lemma: gram derivative L}
	The determinant of the gram matrix $Q = \subscr{A}{21}\subscr{A}{21}^T$ with respect to the elements of $\subscr{A}{21}$ is $\frac{\partial}{\partial \subscr{A}{21}}\det(Q) = 2\det(Q)(Q^{-1}\subscr{A}{21})$
\end{lemma}
\medskip
\begin{proof}
	For $\subscr{A}{21}$, we note the matrix determinant derivative identity
	
	\begin{align*}
	\frac{\partial\det(\subscr{A}{21}B\subscr{A}{21})}{\partial\subscr{A}{21}^T} = 2\det(\subscr{A}{21}B\subscr{A}{21}^T)B\subscr{A}{21}^T(\subscr{A}{21}B\subscr{A}{21}^T)^{-1}.
	\end{align*}
	
	If we set $B = I$, this simplifies to
	
	\begin{align*}
	\frac{\partial\det(Q)}{\partial\subscr{A}{21}^T}= 2\det(Q)\subscr{A}{21}^T(Q)^{-1}.
	\end{align*}
	
	We note that
	
	\begin{align*}
	\frac{\partial\det(Q)}{\partial\subscr{A}{21}} = \left(\frac{\partial\det(Q)}{\partial\subscr{A}{21}^T}\right)^T,
	\end{align*}
	
	\noindent which ultimately yields
	
	\begin{align*}
	\frac{\partial\det(Q)}{\partial\subscr{A}{21}} = 2\det(Q)(Q^{-1}\subscr{A}{21}).
	\end{align*}
\end{proof}

\begin{lemma}{\bf \emph{(Gram Vector Decomposition)}}\label{lemma: gram decomp L}
	For system matrix $\subscr{A}{21}$ with linearly independent rows $\bm{a}_j$, and symmetric positive definite gram matrix $Q = \subscr{A}{21}\subscr{A}{21}^T = PDP^T$ with eigenvalues $\lambda_i$ and eigenvectors $\bm{e}_i$, the total control energy can be represented by $E(\bm{u}) = 12\left( \frac{\sum_{i=1}^M w_i c_i^2}{\sum_{i=1}^M w_i} \right) \sum_{k=1}^M{\frac{1}{\|\bm{a}_k\|^2 \sin(\theta_k)^2}} + \bm{v}_2^T\bm{v}_2$, where $w_i = \prod_{j\neq i}^M \lambda_j$, $c_i = \bm{e}_i^T \bm{v}_1$, and $\theta_i$ is the angle formed by $\bm{a}_k$ and the sub-parallelotope formed by the remaining $\bm{a}_{j \neq k}$.
.\end{lemma}
\medskip
\begin{proof}
	We recall that the total control energy is given by
	
	\begin{align*}
	E_{total} 
	&= 12\bm{v}_1^T Q^{-1}\bm{v}_1 + \bm{v}_2^T\bm{v}_2\\
	&= 12\bm{v}_1^TPD^{-1}P^T\bm{v}_1 + \bm{v}_2^T\bm{v}_2\\
	&= 12(P^T\bm{v}_1)^T D^{-1} (P^T\bm{v}_1) + \bm{v}_2^T\bm{v}_2\\
	&= 12\bm{c}^T D^{-1} \bm{c} + \bm{v}_2^T\bm{v}_2\\
	&= 12\sum_{i=1}^M \frac{c_i^2}{\lambda_i} + \bm{v}_2^T\bm{v}_2
	\end{align*}
	
	We multiply each $k$ term to find a common denominator to yield
	
	\begin{align*}
	E_{total}
	&= 12\frac{\sum_{i=1}^M c_i^2 \prod_{j\neq i}^M \lambda_j}{\prod_j^M \lambda_j} + \bm{v}_2^T\bm{v}_2\\
	&= 12\frac{\sum_{i=1}^M c_i^2 \prod_{j\neq i}^M \lambda_j}{\det(Q)} + \bm{v}_2^T\bm{v}_2\\
	&= 12\left(\frac{\sum_{i=1}^M c_i^2 \prod_{j\neq i}^M \lambda_j}{\sum_{i=1}^M \prod_{j\neq i}^M \lambda_j}\right) \frac{\sum_{k=1}^M \prod_{l\neq k}^M \lambda_l}{\det(Q)}
	\end{align*}
	
	We note that the left term is just a weighted average, with weights $w_i = \prod_{j\neq i}^M \lambda_j$. We also note that $\sum_{k=1}^M \prod_{l\neq i}^M \lambda_l$ is the $M^{th}$ term of the characteristic polynomial of $Q$, which is equivalent to $\sum_{k=1}^M \det(Q_{kk})$, where $Q_{kk}$ represents the $(k,k)$ minor of $Q$. Hence, we write
	
	\begin{align*}
	E_{total}
	&= 12\left(\frac{\sum_{i=1}^M c_i^2 w_i}{\sum_{i=1}^M w_i}\right) \frac{\sum_{k=1}^M \det(Q_{kk})}{\det(Q)}
	\end{align*}
	
	We make use of the geometric fact that the determinant of $Q = \subscr{A}{21}\subscr{A}{21}^T$ is equal to the squared volume of the parallelotope formed by the rows of $\subscr{A}{21}$. We also note that minor $Q_{kk}$ is the gram matrix of $\subscr{A}{21}$ after removing $\bm{a}_k$, represented by $\subscr{A}{21}^{k*}$. Therefore the ratio of the determinants of $Q_{kk}$ and $Q$ becomes the squared ratio of parallelotope volumes with and without $\bm{a}_k$. 
	
	\begin{align*}
	\frac{\det(Q_{kk})}{\det(Q)} = \frac{vol(\subscr{A}{21}^{k*})^2}{vol(\subscr{A}{21})^2}
	\end{align*}
	
	Finally, we realize that the contribution of $\bm{a}_k$ to the parallelotope volume is by a multiple of $\|\bm{a}_k\|\sin(\theta_k)$, where $\theta_k$ is the angle formed by $\bm{a}_k$ and the sub-parallelotope, given by $vol(\subscr{A}{21}) = vol(\subscr{A}{21}^{k*}) \|\bm{a}_k\| \sin(\theta_k)$ to yield
	
	\begin{align*}
	E(\bm{u}) = 12\left( \frac{\sum_{i=1}^M w_i c_i^2}{\sum_{i=1}^M w_i} \right) \sum_{k=1}^M{\frac{1}{\|\bm{a}_k\|^2 \sin(\theta_k)^2}} + \bm{v}_2^T\bm{v}_2
	\end{align*}
	
\end{proof}

\subsection{Validity of the First-Order Approximation}
\label{sec: validation}

Until now, we have derived several useful closed-form expressions from the first-order minimum energy approximation by making the simplifying assumptions \ref{assumption: initial states}, \ref{assumption: network edges}. We explore how well this energy approximation holds when we relax the topological assumption \ref{assumption: network edges}. As a generalized, analytic closed-form energy solution for non-simplified networks is typically intractable or not informative, we compare the first-order energy approximation to a numerical computation of the unsimplified control energy. From linear control theory, we know that for an LTI system $\dot{\bm{x}} = A\bm{x} + B\bm{u}$ obeying the dynamics in \eqref{eq: control dynamics L}, we can define the \textit{Reachability Gramian}:

\begin{align*}
	W_R(0,T) = \int_0^T e^{At}BB^Te^{A^Tt}.
\end{align*}

\noindent The minimum control energy takes the form

\begin{align*}
	E_{NS}(\bm{u}) &= \sum_{i=1}^N \int_0^T u_i(t)^2dt \\
	&= (W_R^{-1}\bm{x}^*)^T W_R (W_R^{-1}\bm{x}^*)\\
	&= \bm{x}^{*T} W_R^{-1}\bm{x}^*,
\end{align*}

\noindent from which we calculate the percent error between the minimum energy of the simplified vs. unsimplified network. The results of this numerical analysis is shown in Fig. \ref{fig:phase}b. 

\end{document}